\documentclass[12pt]{amsart}
\usepackage{amsbsy,amssymb,amscd,amsfonts,latexsym,amstext,delarray,
amsmath,graphicx} 
\input xypic

\usepackage{color}

\newtheorem{thm}{Theorem}[section]
\newtheorem{prop}[thm]{Proposition}
\newtheorem{cor}[thm]{Corollary}
\newtheorem{lem}[thm]{Lemma}

\newtheorem{defn}[thm]{Definition}
\newtheorem{rem}[thm]{Remark}

\numberwithin{equation}{section}

\def\bK{{\mathbb K}}

\def\bS{{\mathbb S}}

\def\C{{\mathbb C}}

\def\N{{\mathbb N}}
\renewcommand{\P}{{\mathbb P}}

\def\R{{\mathbb R}}
\def\Z{{\mathbb Z}}

\def\cA{{\mathcal A}}

\def\cC{{\mathcal C}}
\def\cD{{\mathcal D}}

\def\cF{{\mathcal F}}
\def\cG{{\mathcal G}}
\def\cH{{\mathcal H}}
\def\cI{{\mathcal I}}
\def\cJ{{\mathcal J}}

\def\cM{{\mathcal M}}
\def\cN{{\mathcal N}}

\def\cP{{\mathcal P}}
\def\cQ{{\mathcal Q}}

\def\cS{{\mathcal S}}
\def\cT{{\mathcal T}}
\def\cU{{\mathcal U}}
\def\cV{{\mathcal V}}

\def\cZ{{\mathcal Z}}

\def\Ker{{\rm Ker}}

\def\Sp{{\rm Spec}}

\def\Tr{{\rm Tr}}

\def\fC{{\mathfrak C}}

\title[Gamma Spaces and Information]{Gamma Spaces and Information} 
\author{Matilde Marcolli}
\date{2018}
\address{California Institute of Technology \\ USA \newline \indent
Perimeter Institute for Theoretical Physics \\ Canada \newline \indent
University of Toronto \\ Canada}
\email{matilde@caltech.edu}
\email{matilde@math.utoronto.ca}
\email{mmarcolli@perimeterinstitute.ca}

\begin{document}
\maketitle

\begin{abstract}
We investigate the role of Segal's Gamma-spaces in the context of
classical and quantum information, based on categories of finite probabilities
with stochastic maps and density matrices with quantum channels. 
The information loss functional extends to the setting of probabilistic
Gamma-spaces considered here. The Segal construction of connective spectra 
from Gamma-spaces can be used in this setting to obtain spectra 
associated to certain categories of gapped systems.
\end{abstract}

\tableofcontents

\section{Introduction}

Segal introduced $\Gamma$-spaces in \cite{Segal}, as a way of constructing
connective spectra from the data of a category $\cC$ with a zero object and a categorical sum.
A $\Gamma$-space is a functor $F_\cC: \Gamma^0 \to \Delta_*$ from the category
of pointed finite sets to the category of pointed simplicial sets. Spectra associated to
$\Gamma$-spaces are obtained by extending the functor to an endofunctor of $\Delta_*$
and applying it to the spheres $S^n$, so that one obtains a spectrum given by the sequence
of spaces $X_n=F_\cC(S^n)$ with structure maps
$S^1\wedge F_\cC(S^n) \to F_\cC(S^1\wedge S^n)=X_{n+1}$. In particular, when $F_\cC: \Gamma_0\hookrightarrow \Delta_*$ is the natural inclusion, that is, 
the $\Gamma$-space associated to the category $\cC=\Gamma^0$ itself, one obtains the
sphere spectrum. More generally, 
it is known that all connective spectra can in fact be obtained in this way, \cite{BousFried}, 
\cite{Segal}, \cite{Thoma}. Moreover, $\Gamma$-spaces provide a very transparent
description of the smash product of spectra, \cite{Lyd}. 

\smallskip

Throughout this paper we will use equivalently the $\Gamma$ notation of Segal 
for the category of finite pointed sets as well as the notation $\cS_*$. We trust that
this will not be a source of confusion. 

\smallskip

The main purpose of this paper is developing probabilistic versions of the Segal construction, 
in the setting of classical and quantum information. In the classical setting, we
replace the usual category of finite pointed sets used in homotopy theoretic
construction with a probabilistic version. Heuristically this corresponds to
finite sets where each of the points can be the base point with certain
assigned probabilities. The more precise formulation is in terms of a
form of wreath product between the category of finite pointed sets and
the category of finite probabilities with stochastic maps. More generally,
one can start with any category $\cC$ with a zero object and a categorical
sum (coproduct) and form a similar wreath product with the category $\cF\cP$
of finite probabilities. The resulting category $\cP\cC$ still has a zero object and a
coproduct, where on the probabilities the coproduct is a product of
statistically independent distributions. The Segal construction \cite{Segal}
of $\Gamma$-spaces from categories with zero object and sum can then
be applied to the categories $\cP\cC$ leading to associated spectra.
For the purpose of obtaining an explicit description 
in the particular case where $\cC$ is the category $\cS_*$ of finite pointed
sets, we reformulate the usual notion of $\Gamma$-spaces using the
category of cubical sets $\Box_*$ instead of simplicial sets. We then analyze
the explicit form of the $\Gamma$-space $F_{\cP\cS_*}: \Gamma^0 \to \Box_*$. 
We then show that the notion of $\Gamma$-space 
itself can be made probabilistic, by considering 
functors $F: \cP\cS_* \to \cP\Box_*$ between the corresponding
probabilistic categories. These map a probabilistic pointed set $\Lambda X$
to a probabilistic pointed cubical set given by a $\Lambda$-convex combination
of the cubical sets associated to the pointed sets in $\Lambda X$ by
the original (non-probabilistic) $\Gamma$-space. 

\smallskip

An information loss functional of the kind considered in \cite{BFL}
can be defined on the category of finite probabilities $\cF\cP$
and extended to the categories $\cP\cS_*$ and $\cP\Box_*$
of probabilistic pointed sets and probabilistic pointed
cubical sets. As in the case of \cite{BFL} these information
loss functionals are determined by a set of Khinchin-type
axioms and are always expressible in terms of a difference of
Shannon entropies and of topological invariants of cubical
(or simplicial) sets. 

\smallskip

We consider then a quantum information version $\cQ\cC$ of
the probabilistic categories $\cP\cC$, again starting with
a category $\cC$ with zero object and sum. In this case
we consider a category $\cF\cQ$ of finite quantum probabilities
(density matrices) with morphisms given by quantum channels.
 The objects of the category $\cQ\cC$ are collections of
 pairs of objects in $\cC$ with an assigned amount of quantum interference 
 (coherence) specified by the
 entries of a density matrix. A similar version $\cQ\cA\cC$
 uses a category $\cA\cC$ of arrows of $\cC$ with associated
 density matrices. The category $\cQ\cC$ still has a zero
 object and coproduct, hence one can again apply
 the Segal construction and obtain $\Gamma$-spaces
 (as well as corresponding probabilistic $\Gamma$-spaces)
 and associated spectra. In the case where the underlying
 deterministic category is $\cC=\cS_*$ the category of
 finite pointed sets, one can again describe explicitly the
 resulting topological spaces, from which one can
 see directly that they are topologically more complex
 than their counterparts in the case of classical probabilities. 
 
 \smallskip
 
 Finally, we consider a category $\cF\cQ^\Delta$ of
 gapped systems related by gap preserving quantum
 channels, and corresponding categories $\cQ\cC^\Delta$
 realized as subcategories of $\cQ\cC$.  We compute explicltly
 the associated $\Gamma$-space in the case of $\cC=\cS_*$.
 These $\Gamma$-spaces provide a natural construction of
 connective spectra associated to gapped systems.
 
 \smallskip
 
 In a broader perspective, we regard this as a first step in the
 direction of developing a {\em probabilistic version of homotopy theory}.
 There are several reasons why this would be a desirable goal.
 Mainly, one can think of the following sources of motivation:
 \begin{itemize}
 \item {\em Physics}: as mentioned above spectra and other homotopy theoretic 
 methods are widely regarded as an important approach to study gapped systems and
 topological phases of matter, \cite{Gaio}, \cite{Xiong}. 
 Because of the quantum information formulation
 of the problem, it is expected that a development of homotopy theoretic
 methods that incorporate classical and quantum probabilities will be
 a useful tool in this setting.
 \item {\em Statistics}: data analysis has incorporated topological methods
 in the form of persistent topology and persistent homology, which are based
 on simplicial sets associated to a collection of data points, \cite{Carlsson}. A version of simplicial sets
 and homotopy theoretic methods that incorporate probabilities may be useful
 in developing better categorical structures in persistent topology 
 (model categories, etc.). This question was suggested by Jack Morava. 
 \item {\em Neuroscience}: the idea of enriching algebraic structures with
 probabilities is not new, and was in fact developed in great generality in \cite{Grenander}
 and found a wide range of applications, including ``pattern theory" in models of
 computer vision and neuroscience. The use of simplicial and homotopy theoretic
 methods in neuroscience is also advocated in \cite{Curto}, \cite{Manin} and one
 expects that an adequate treatment will have to include probabilities along
 with combinatorics, \cite{MarTsao}.
  \end{itemize}
While the present paper only focuses on one particular aspect, namely
probabilistic versions of the Segal construction of spectra through $\Gamma$-spaces,
and does not touch upon these broader motivations and applications, one should
regard the purpose of this investigation within this more general context. 

\medskip
\section{From pointed sets to probabilities}

The category of pointed finite sets is a category with a zero object and a categorical sum,
hence one can apply to it the construction of \cite{Segal} that produces a $\Gamma$-space
and a spectrum, which in this case is the sphere spectrum $\bS=\{ S^n \}$, a result known as the
Barratt--Priddy--Quillen theorem. 

\smallskip

The main heuristic observation that we want to formalize in this paper is the fact that
the Segal notion of $\Gamma$-spaces relies crucially on the construction of a category
of {\em summing functors} (which we will review briefly). These behave
very much like measures, so that one is lead to believe that the use of pointed sets
in the construction should in fact be seen as a proxy for a measure theoretic setting.
The first step in making this heuristic observation more rigorous is to consider
generalizations of the category of pointed sets based on finite probability spaces.
The main idea is to think of pointed sets as a special case where the probability
is a delta function supported at the base point and morphisms are measure preserving maps 
(hence pointed maps).

\smallskip

The category of finite sets with probability measures considered in \cite{BFL}, \cite{BF} 
with measure preserving morphisms satisfying \eqref{measpres} is not directly
suitable for the application of the Segal construction of \cite{Segal}. Thus, we
consider here some possible modifications, in the form of categories of finite sets 
with probability distributions and with stochastic maps.

\smallskip
\subsection{Finite probabilities and stochastic map}

We consider a category of finite probability spaces and
stochastic matrices defined as follows.

\begin{defn}\label{finprobS}
Let $\cF\cP$ be the category whose objects are pairs $(X,P)$ of a 
finite set $X$ with a probability measure $P$. Morphisms 
$$S\in {\rm Mor}_{\cF\cP}((X,P),(Y,Q))$$ are stochastic $(\# Y \times \# X)$-matrices 
$S$, with the following properties:
\begin{enumerate}
\item $S_{yx}\geq 0$, for all $x\in X$, $y\in Y$;
\item $\sum_{y\in Y} S_{yx}=1$ for all $x\in X$;
\item the probability measures are related by $Q = S\, P$.
\end{enumerate}
\end{defn}

\begin{rem}\label{FinStoch}{\rm 
This category of stochastic matrices is not the same as the
category {\rm FinStoch} of \cite{Fritz}, where objects are finite
sets without measures, so that the third condition of 
Definition~\ref{finprobS} above is not
required. However, $\cF\cP$ is the under category 
$1/{\rm FinStoch}$, as discussed after Definition~3 of \cite{BF}.}
\end{rem}

\smallskip

Note that to a stochastic matrix $S$ as above we can associate a multivalued
function $f_S: X \to Y$ with $f_S(x_j)=\{ y_i \in Y\,:\, S_{ij} >0 \}$. The morphisms
are measure preserving in the sense that the relation
\begin{equation}\label{Smeaspres}
 Q_y = \sum_{x\in X} S_{yx} P_x 
\end{equation}
holds, replacing the original \eqref{measpres}. The category of finite probability
measures considered in \cite{BFL} is a subcategory of $\cF\cP$ consisting 
of those morphisms where, for each $x$ there is a unique $y=y(x)$ such that $S_{yx}>0$.
In this case, by the stochastic condition this value must be $S_{yx}=1$ hence $S$
corresponds to a (single valued) function $f: X \to Y$ satisfying \eqref{Smeaspres},
which in this case becomes the same as \eqref{measpres},
$$ Q_y =\sum_{x \in f^{-1}(y)} P_x . $$
The sets of morphisms  ${\rm Mor}_{\cF\cP}((X,P),(Y,Q))$ are convex sets.

\smallskip

\begin{rem}\label{pointed}
The usual category of finite pointed sets can be seen as the subcategory
of $\cF\cP$ given by elements of the form $(X,\delta_{x_0})$ and
morphisms given by stochastic maps $S: (X,\delta_{x_0}) \to (Y,\delta_{y_0})$
of the form $S_{yx}=\chi_{f^{-1}(y)}(x)$, with $\chi$ the indicator function,
so that $\delta_{y_0,y}=\sum_x S_{yx} \delta_{x_0,x}=
\sum_{x\in f^{-1}(y)} \delta_{x_0,x}$.
\end{rem}

\smallskip

\begin{rem}\label{fuzzy}
The objects $(X,P)$ of the category $\cF\cP$ can be thought of
as fuzzy sets, with $P_x$ the value at the point $x\in X$
of the membership function of the fuzzy set, \cite{Zadeh}.
\end{rem}

\smallskip
\subsubsection{Zero object} The singletons in the category of finite
probabilities are zero objects.

\begin{lem}\label{0obcatsum}
The category $\cF\cP$ has zero objects given by singleton sets
$(\{ x \},1)$. 
\end{lem}

\proof For all objects $(Y,Q)$ there is a unique morphism $\hat Q: (\{ x \},1) \to (Y,Q)$
which is given by $\hat Q_{yx}=Q_y$, and a unique morphism $\hat 1: (Y,Q) \to (\{x\},1)$
given by $\hat 1_{xy}=1$ for all $y\in Y$, so that $1=\sum_{y\in Y} Q_y =\sum_{y\in Y} \hat 1_{xy} Q_y$. 
Thus, $(\{ x \},1)$ is a zero-object. 
\endproof

\smallskip

We will use the notation $\hat 1_{(X,P)}$ for the unique morphism
$\hat 1_{(X,P)}: (X,P) \to (\{x\},1)$, whenever it is useful to keep track explicitly
of the source object $(X,P)$.

\smallskip
\subsubsection{The target morphism}

Given any pair of objects $(X,P)$ and $(Y,Q)$ in $\cF\cP$, there is always a
distinguished morphism, which we denote by $\hat Q: (X,P) \to (Y,Q)$, which is
defined by $\hat Q_{ba}=Q_b$. This clearly satisfies $Q_b=\sum_a \hat Q_{ba} P_a$.
We refer to this morphism as ``the target morphism".

\smallskip

\begin{lem}\label{targetmor}
The target morphism $\hat Q: (X,P) \to (Y,Q)$ has the property that, given any morphism
$S: (X',P')\to (X,P)$ and any morphism $S': (Y,Q) \to (Y',Q')$ the compositions satisfy
$\hat Q \circ S = \hat Q$ and $S'\circ \hat Q=\hat Q'$.
\end{lem}

\proof We have $(\hat Q \circ S)_{ba'}=\sum_a \hat Q_{ba} S_{aa'} = Q_b \sum_a S_{aa'}=Q_b$ for all $a'\in X'$,
and $(S'\circ \hat Q)_{b'a} = \sum_{a'} S'_{b'a'} Q_{a'} =Q'_{b'}$ for all $a\in X$.
\endproof

\smallskip

\begin{rem}\label{zeromor}{\rm 
The target morphisms are the categorical zero morphisms, that is, 
the morphisms that factor through the zero object. }
\end{rem}

\smallskip
\subsection{Coproduct of finite probabilities}

We want to construct a coproduct of finite probabilities which reduces to the coproduct
of pointed sets in the case where the measures are delta measures.

\smallskip
\subsubsection{Finite probabilities as combinations of pointed sets}

We can equivalently regard a finite probability $(X,P)$ as a finite set $X$ where
each point $x\in X$ can be chosen as the base point with probability $P_x$. Thus,
we can regard the object $(X,P)$ as a formal convex combination of pointed sets,
\begin{equation}\label{XPcomb}
(X,P) = \sum_{x\in X} P_x \, (X, x) .
\end{equation}

This interpretation means that we can embed the category $\cF\cP$ of
finite probabilities in a category $\cP\cS_*$ of probabilistic pointed sets
defined as follows.

\smallskip

\begin{defn}\label{combptset}
The category $\cP\cS_*$ of probabilistic pointed sets has objects that are convex combinations
of pointed sets 
$$ \Lambda X = \sum_i \lambda_i (X_i,x_i), $$
where $\Lambda=(\lambda_i)$ wtih $\lambda_i\geq 0$ and $\sum_i \lambda_i=1$ and $X=\{ (X_i,x_i) \}$
a finite collection of pointed sets. The morphisms in $\cP\cS_*$ are given by
$\Phi \in {\rm Mor}_{\cP\cS_*}(\Lambda X, \Lambda' X')$ consisting of a pair $\Phi=(S,F)$
\begin{enumerate}
\item $S$ is a stochastic map with $S\Lambda =\Lambda'$
\item $F=(F_{ji})$ is a collection of probabilistic pointed maps $F_{ji}: (X_i,x_i) \to (X'_j,x'_j)$.
\end{enumerate}
Here a probabilistic pointed map $F_{ji}$ is a finite set $\{ F_{ji,a} \}$ of pointed maps
$F_{ji,a}: (X_i,x_i) \to (X'_j,x'_j)$ together with a set of probabilities $\mu^{(ji)}_a$ with
$\sum_a\mu^{(ji)}_a=S_{ji}$.
\end{defn}

\smallskip

Thus we regard a morphism between two probabilistic pointed sets as a collection of
probabilistic pointed maps $F=\{ F_{ji,a} \}$. This means that for a fixed source $(X_i,x_i)$ 
and a point $x\in (X_i,x_i)$ the value $F(x)$ is obtained by first choosing a map
$F_{ji,a}$ with probability $\mu^{(ji)}_a$. Equivalently, one chooses the set 
$F_{ji}=\{ F_{ji,a} \}$ with probability $S_{ji}$, which is the sum of all the probabilities
$\mu^{(ji)}_a$ of choosing one of the maps $F_{ji,a}$ in the set.
This includes the case where $F_{ji}$ consists of a single pointed map applied 
with probability $S_{ji}$. 

\smallskip

\begin{rem}\label{composePhi}{\rm
The composition of two morphisms $\Phi=(S,F): \Lambda X \to \Lambda' X'$ and
$\Phi'=(S',F'): \Lambda' X' \to \Sigma Y$ is given by $\Phi'\circ \Phi =(S' \circ S,  F'\circ F)$,
where $S'\circ S$ is the product of the stochastic matrices and $F'\circ F =\{ (F'\circ F)_{ki} \}$ with
the set $(F'\circ F)_{ki}=\{ F'_{kj,a}\circ F_{ji,b} \}$ with probabilities $\mu^{(kj)}_a \mu^{(ji)}_b$
with $\sum_{a,b,j} \mu^{(kj)}_a \mu^{(ji)}_b =\sum_j S'_{kj} S_{ji} =(S'\circ S)_{ki}$, so that
the probability associated to the set $(F'\circ F)_{ki}$ in $F'\circ F$ is $(S'\circ S)_{ki}$.}
\end{rem}

\smallskip

\begin{rem}\label{embedFPSP} {\rm
An embedding of the category $\cF\cP$ in the category $\cP\cS_*$ is obtained
by mapping $\Lambda=(\lambda_i)$ to the set $\Lambda \star =\sum_i \lambda_i (\{ \star_i \}, \star_i)$.
and morphisms $S \Lambda =\Lambda'$ to $\Phi=(S, {\bf 1})$ with ${\bf 1}=\{ 1_{ji} \}$ with
probabilities $S_{ji}$. }
\end{rem}

\smallskip

\begin{rem}\label{forget}{\rm 
There is a forgetful functor from $\cP\cS_*$ to $\cF\cP$ that maps
$\Lambda X$ to the finite probability $\Lambda$ and a
morphism $\Phi=(S,F)$ to the stochastic matrix $S$. }
\end{rem}

\smallskip
\subsubsection{Zero objects}
The category $\cP\cS_*$ of probabilistic pointes sets also has zero objects
given by singletons.

\begin{lem}\label{zeroobjPS}
The objects $\Lambda X$ given by a singleton set $X=(\{ x\}, x)$ and $\Lambda=1$
are zero objects in $\cP\cS_*$.
\end{lem}

\proof Given any object $\Lambda X=\sum_i \lambda_i (X_i,x_i)$, there is a unique
morphism $\Phi=(S,F): (\{ x\}, x) \to \sum_i \lambda_i (X_i,x_i)$ with $S=\hat\Lambda$,
the unique morphism $\hat\Lambda$ from the zero-obejct $(\{ x\},1)$ of $\cF\cP$ to
the finite probability $\Lambda$ and with $F=(F_i)$ with $F_i: x \mapsto x_i$ with 
probability $\lambda_i$. Moreover,
there is also a unique morphism $\Phi=(S,F): \Lambda X \to (\{ x\}, x)$ where $S=\hat 1_{\Lambda}$
is the unique morphism in $\cF\cP$ from the finite probability $\Lambda$ to the zero
object $(\{ x\},1)$ and $F=(F_i)$ with $F_i: (X_i,x_i)\to (\{ x\}, x)$ the constant function
with probability $1$.
\endproof

\smallskip
\subsubsection{Coproduct of probabilistic pointed sets}

The category $\cP\cS_*$ has a coproduct inherited from the category of pointed sets.

\begin{defn}\label{defcoprodSP}
Given $\Lambda X=\sum_{i=1}^N (X_i,x_i)$ and $\Lambda' X'=\sum_{j=1}^M \lambda'_j (X'_j,x'_j)$
we have
\begin{equation}\label{coprodSP}
\Lambda X \amalg \Lambda' X' := \sum_{ij} \lambda_i \lambda'_j \,\, (X_i,x_i)\vee (Y_j,x'_j), 
\end{equation}
with the usual coproduct of pointed sets
$$ (X_i,x_i)\vee (Y_j,x'_j) = (X_i \sqcup Y_j / x_i\sim y_j , x_i\sim y_j) . $$
\end{defn}

\smallskip

In the case of two probabilistic pointed sets obtained from two finite probability
distributions as in \eqref{XPcomb}, this corresponds to the intuition that 
one considers each point $x\in X$
as the base point with probability $P_x$, and similarly with $X'$. Thus,
when forming the coproduct, the probability that it is obtained by
identifying $x\in X$ with $x'\in X'$ is the product of the probabilities
$P_x$ and $P'_{x'}$, namely the probability of independently choosing
$x$ and $x'$ as the respective base points. 

\smallskip

\begin{lem}\label{prodcoprod}
The coproduct induced by \eqref{coprodSP} on probabilistic pointed sets 
obtained from finite probability distributions as in \eqref{XPcomb} is 
the product of statistically independent probabilities
\begin{equation}\label{coprodprob}
(X,P)\amalg (X',P') = (X\times X', P\cdot P').
\end{equation}
\end{lem}

\proof In the case of two probabilistic pointed sets obtained from two finite probability
distributions as in \eqref{XPcomb}, this coproduct is given by
\begin{equation}\label{coprodprob2}
(X,P)\amalg (X',P') = \sum_{x,x'} P_x P'_{x'} \, (X,x)\vee (X',x').
\end{equation}
We can interpret the right-hand-side
of \eqref{coprodprob} as a probability space by applying the forgetful functor
from $\cP\cS_*$ to $\cF\cP$. This identifies it with the finite probability
$(X\times X', P\cdot P')$.
\endproof

\smallskip

\begin{rem}\label{forgetX}{\rm
It is preferable to work with the category $\cP\cS_*$ rather than with $\cF\cP$, since
just retaining the information of the finite probability would give the same product probability 
space for any underlying binary operation on probabilistic pointed sets, without remembering 
the specific operation on the underlying sets.
}\end{rem}

\smallskip
\subsubsection{Statistical independence: product or coproduct?}\label{prodcoprodSec}

It is well known that the category of measure spaces or of finite
probability spaces with measure preserving map (hence also 
the category $\cF\cP$ considered here) does not have 
a universal categorical product, that is, an object $X_1\times X_2$
such that, for all morphisms $f_1: X \to X_1$ and $f_2: X \to X_2$
there exists a unique morphism $h: X \to X_1 \times X_2$ such
that the diagram commutes
\begin{equation}\label{proddiagr}
 \xymatrix{ & X \ar[dl]_{f_1} \ar[dr]^{f_2} \ar[d]^h & \\
X_1 & X_1 \times X_2 \ar[l]_{\pi_1} \ar[r]^{\pi_2} & X_2
} \end{equation}

\smallskip

However, as shown in \cite{Franz}, the category of finite
probability measures with measure preserving maps is
a ``tensor category with projections", namely a tensor
category $(\cC,\otimes)$ together with two natural
transformations $\pi_i: \otimes \to \Pi_i$ where
$\Pi_i(X_1,X_2)=X_i$, such that for any pairs of morphisms
$f_i: Y_i \to X_i$, the diagram commutes:
$$ \xymatrix{  Y_1 \ar[d]^{f_1} & Y_1\otimes Y_2 \ar[l]_{\pi_{Y_1}} \ar[r]^{\pi_{Y_2}} \ar[d]^{f_1\otimes f_2} & Y_2 \ar[d]^{f_2} \\
X_1 & X_1\otimes X_2 \ar[l]_{\pi_{X_1}} \ar[r]^{\pi_{X_2}} & X_2
} $$
In a tensor category with projections two morphisms $f_i: X \to X_i$ are {\em independent}
if there exists a morphism $h: X \to X_1 \otimes X_2$ such that the product diagram 
\eqref{proddiagr} commutes. In the category of finite probabilities with measure
preserving maps this notion of independence agrees with the usual notion
of stochastic independence, with 
$$ (X_1,P_1)\otimes (X_2, P_2)=(X_1\times X_2, P_1 P_2) $$
the product of independent probability spaces. 

\smallskip

Tensor categories with projections (semicartesian
monoidal categories) have the property that 
projections are unique when they exist, and their existence is
equivalent to the terminality of the unit, \cite{Leinster2}.

\smallskip

The point of view discussed here shows that the product of
statistically independent measures may be interpreted as
a coproduct instead of a product on the category $\cF\cP$
of finite measures.

\smallskip
\subsection{Universal property of the coproduct of probabilistic pointed sets} 

The coproduct of Definition~\ref{defcoprodSP} in the category $\cP\cS_*$ satisfies the
universal property.

\smallskip

\begin{defn}\label{FveeF}
Given objects $\Lambda X =\sum_a \lambda_a (X_a,x_a)$ and $\Sigma Y=\sum_k \sigma_k (Y_k,y_k)$
and morphisms $\Phi=(S,F): \Lambda X \to \Sigma Y$ 
and $\Phi'=(S',F'): \Lambda' X' \to \Sigma Y$ with $S\Lambda =\Sigma$ and $S'\Lambda'=\Sigma$
and with $F=\{ f_{ka,r} \}_{r=1}^N$
with probabilities $\sum_r \mu^{(ka)}_r =S_{ka}$ and $F'=\{ f'_{ka',r'} \}_{r'=1}^M$ 
with probabilities $\sum_{r'} \mu^{(ka)}_{r'} =S'_{ka}$ we define 
$F\vee F'$ as the collection $\{ f_{ka,r} \vee f'_{ka',r'} \}$ of pointed maps from the 
coproducts of pointed sets with probabilities $\sigma_k^{-1} \mu^{(ka)}_r \mu^{(ka')}_{r'}$
for $\sigma_k\neq 0$ and $M^{-1} \mu^{(ka)}_r + N^{-1} \mu^{(ka')}_{r'}$ for $\sigma_k=0$.
\end{defn}

\smallskip

\begin{thm}\label{coproddiagr}
For any objects $\Lambda X$ and $\Lambda' X'$ in $\cP\cS_*$, 
there are unique morphisms $\Psi: \Lambda X \to \Lambda X \amalg \Lambda' X'$
and $\Psi': \Lambda' X' \to \Lambda X \amalg \Lambda' X'$ such that,
for any object $\Sigma Y$, with $\Sigma=(\sigma_k)$, in $\cP\cS_*$ and any choice of
morphisms $\Phi=(S,F): \Lambda X \to \Sigma Y$
and $\Phi'=(S',F'): \Lambda' X' \to \Sigma Y$, there is a unique
morphism $\Phi\amalg \Phi': \Lambda X\amalg  \Lambda' X'  \to \Sigma Y$ such that
the diagram commutes
\begin{equation}\label{univdiagFP0}
\xymatrix{  & \Sigma Y & \\
\Lambda X \ar[ur]^{\Phi} \ar[r]^{\Psi \quad} & \Lambda X\amalg  \Lambda' X' 
\ar[u]_{\Phi\amalg \Phi'} &
\Lambda' X' \,. \ar[ul]_{\Phi'} \ar[l]_{\quad \Psi'} 
} \end{equation}
where $(\Phi\amalg_\lambda \Phi') =(S\amalg S', F\vee F')$ with 
\begin{equation}\label{coprodmor}
(S\amalg_\lambda S')_{k,(a,a')} =\left\{ \begin{array}{ll} 
\sigma_k^{-1} \cdot S_{k,a} \cdot S'_{k,a'}   & \sigma_k \neq 0 \\[3mm]
S_{k,a} + S'_{k,a'} & \sigma_k =0.
\end{array}\right. 
\end{equation}
and with $F\vee F'$ as in Definition~\ref{FveeF}.
\end{thm}

\proof We have $\Psi=(\cI, \cF)$ and $\Psi'=(\cI',\cF')$ where
the morphisms $\cI\in {\rm Mor}_{\cF\cP}(\Lambda, \Lambda \cdot \Lambda')$,
$\cI'\in {\rm Mor}_{\cF\cP}(\Lambda', \Lambda \cdot \Lambda')$ are given by
$$ (\cI)_{(b,b'),a} = \delta_{ab} \, \lambda'_{b'} \ \ \ \text{ and } \ \ \ 
(\cI')_{(b,b'),a'} = \delta_{a'b'} \, \lambda_b. $$
These satisfy $\cI \Lambda = \Lambda \cdot \Lambda'$ and
$\cI' \Lambda' = \Lambda \cdot \Lambda'$. 
The probabilistic pointed maps 
$\cF=(\cF_{(b,b'),a})$ and $\cF'=(\cF'_{(b,b'),a'})$ are given by the standard pointed
inclusion maps to the coproduct of pointed sets $\cF_{(b,b'),a}=\delta_{ab} \cF_{b'b}$
with $\cF_{b'b}: (X_b,x_b) \hookrightarrow (X_b,x_b)\vee (X'_{b'},x'_{b'})$ the inclusion map
of the coproduct of pointed sets, chosen with probability $\lambda'_{b'}$, and 
similarly $\cF'_{(b,b'),a'}=\delta_{a' b'} \cF'_{bb'}$ with
and $\cF'_{bb'}: (X'_{b'},x'_{b'}) \hookrightarrow (X_b,x_b)\vee (X'_{b'},x'_{b'})$ the inclusions
taken with probabilities $\lambda_b$.

\smallskip

A morphism $\Theta: \Lambda X\amalg  \Lambda' X'  \to \Sigma Y$ is given by 
$\Theta=(\tilde S, \tilde F)$ where $\tilde S \in {\rm Mor}_{\cF\cP} (\Lambda \cdot \Lambda', \Sigma)$
and $\tilde F=\{ \tilde F_{k,(i,j)} \}$ is a collection of pointed maps $\tilde F_{k,(i,j)}=\{ \tilde F_{k,(i,j),\, s} \}$
from the pointed sets $(X_i,x_i)\vee (X'_j,x'_j)$ to the pointed set $(Y_k,y_k)$, with
probabilities $\sum_s \mu^{(k,(i,j))}_s =\tilde S_{k, (i,j)}$.

\smallskip

The compositions
$\Theta \circ \Psi$ and $\Theta \circ \Psi'$ are given on the stochastic matrices by 
\begin{equation}\label{tildeS}
\sum_{a,a'} \tilde S_{k,(a,a')} \lambda_a \lambda'_{a'}= \sigma_k, \ \ \ \text{ for } \ \ \Sigma=(\sigma_k)
\end{equation}
while the compositions with $\cI$ and $\cI'$ are given by 
\begin{equation}\label{comp1}
\sum_{a,a'} \tilde S_{k,(a,a')} \cI_{(a,a'),b}=
\sum_{a,a'} \tilde S_{k,(a,a')} \delta_{ab} \lambda'_{a'} = \sum_{a'} \tilde S_{k,(b,a')} \lambda'_{a'}
\end{equation}
\begin{equation}\label{comp2}
\sum_{a,a'} \tilde S_{k,(a,a')} \cI'_{(a,a'),b'}=\sum_a \tilde S_{k,(a,b')} \lambda_a.
\end{equation}
At the level of the pointed maps we have the 
compositions $\tilde F \circ \cF$ and $\tilde F\circ \cF'$, which are compositions
with the inclusions $\tilde F \circ \cF=\{ \tilde F_{k,(i,j),s} \circ \cF_{ji} \}$ with
probabilities $\mu^{(k, (i,j))}_s \lambda'_j$ and 
$\tilde F\circ \cF'=\{ \tilde F_{k,(i,j),s} \circ \cF'_{ij} \}$, with probabilities
$\mu^{(k, (i,j))}_s \lambda_i$. 

\smallskip

The morphism $\tilde S_{k,(a,a')}=\sigma_k^{-1} S_{k,a} S'_{k,a'}$ for $\sigma_k\neq 0$
and $\tilde S_{k,(a,a')}=S_{k,a} + S'_{k,a'}$ when $\sigma_k =0$ satisfies \eqref{tildeS},
since in the case $\sigma_k\neq 0$
$$ \sum_{a,a'} \sigma_k^{-1} S_{k,a} S'_{k,a'} \lambda_a \lambda'_{a'}= \sigma_k^{-1} \sigma_k^2 $$
while in the case with $\sigma_k =0$ we have 
$\sum_{a,a'} (S_{k,a} + S'_{k,a'}) \lambda_a \lambda'_{a'}= 0$ since $\sum_a S_{k,a}\lambda_a =
\sum_{a'} S'_{k,a'} \lambda'_{a'}= 0$.  Moreover, we have
\begin{equation}\label{comp3}
\begin{array}{ll}
 \sum_{a,a'} \sigma_k^{-1} S_{k,a} S'_{k,a'} \cI_{(a,a'),b} = \sigma_k^{-1} S_{k,b} \sum_{a'} S'_{k,a'}\lambda_{a'} = S_{k,b} &  \sigma_k\neq 0 \\[3mm] 
 \sum_{a,a'}  (S_{k,a} + S'_{k,a'}) \cI_{(a,a'),b} = S_{k,b} + \sum_{a'}S'_{k,a'}\lambda_{a'} =S_{k,b} & \sigma_k=0.
\end{array}
\end{equation}

\smallskip

By the universal property of the coproduct of pointed sets, there is a unique map
$f_{ki,r}\vee f'_{kj,r'}$ with the property that $(f_{ki,r}\vee f'_{kj,r'})\circ \cF_{ji}=f_{ki,r}$
and $(f_{ki,r}\vee f'_{kj,r'})\circ \cF'_{ij}=f'_{kj,r'}$. 
We consider the resulting map $F\vee F'$ as in Definition~\ref{FveeF}, with the probability 
assigned to $f_{ki,r}\vee f'_{kj,r'}$ given by $\sigma_k^{-1} \mu^{(ki)}_r \mu^{(kj)}_{r'}$
for $\sigma_k\neq 0$ and $\mu^{(ki)}_r + \mu^{(kj)}_{r'}$ for $\sigma_k=0$. Then
the probability associated to the composition $(f_{ki,r}\vee f'_{kj,r'})\circ \cF_{ji}$ is 
given by $\sigma_k^{-1} \mu^{(ki)}_r \mu^{(kj)}_{r'} \lambda'_j$ when $\sigma_k\neq 0$
and $(M^{-1} \mu^{(ki)}_r + N^{-1}\mu^{(kj)}_{r'}) \lambda'_j$ when $\sigma_k =0$. Since for all
$j,r'$ the composition is equal to $f_{ki,r}$ the probabilities correspondingly should add up.
Indeed, we have
$$ \sum_{j,r'} \sigma_k^{-1} \mu^{(ki)}_r \mu^{(kj)}_{r'} \lambda'_j =
\mu^{(ki)}_r \sigma_k^{-1} \sum_j S'_{kj} \lambda'_j =\mu^{(ki)}_r  $$
when $\sigma_k\neq 0$, while in the 
case with $\sigma_k=0$ we have 
$$ \sum_{j,r'} (M^{-1} \mu^{(ki)}_r + N^{-1} \mu^{(kj)}_{r'}) \lambda'_j =\mu^{(ki)}_r + N^{-1} \sum_j 
S'_{kj}\lambda'_j = \mu^{ki}_r . $$
The counting of probabilities for the compositions $(f_{ki,r}\vee f'_{kj,r'})\circ \cF'_{ij}=f'_{kj,r'}$
is analogous. Thus, we find that there is a unique choice of $\tilde F=F \vee F'$ with the property 
that $\tilde F\circ \cF=F$ and $\tilde F\circ \cF'=F'$. 
Thus, we obtain that $(\Phi\amalg_\lambda \Phi')\circ \Psi=\Phi$ and 
$(\Phi\amalg_\lambda \Phi')\circ \Psi'=\Phi'$.
\endproof

\smallskip
\subsection{Probabilistic categories as wreath products}

The same procedure we used to pass from the category of finite pointed sets $\cS_*$ to
its probabilistic counterpart $\cP\cS_*$ can be generalized to a procedure that associates
to a category $\cC$ with a zero object $0$ and a categorical sum (coproduct) $\oplus$ a
new category $\cP\cC$, which is the probabilistic version of $\cC$.

\smallskip

\begin{defn}\label{catPC}
 $\cP\cC$ is the category whose objects are formal finite convex combinations
$$ \Lambda C = \sum_i \lambda_i C_i, $$
with $\Lambda=(\lambda_i)$ with $\sum_i \lambda_i =1$ and $C_i \in {\rm Obj}(\cC)$
and with morphisms $\Phi: \Lambda C \to \Lambda' C'$ given by pairs $\Phi=(S,F)$
with $S$ a stochastic matrix with $S\Lambda =\Lambda'$ and $F=\{ F_{ab,r} \}$
a finite collection of morphisms $F_{ab,r}: C_b \to C'_a$ with assigned probabilities 
$\mu^{ab}_r$ with $\sum_r \mu^{ab}_r =S_{ab}$. 
\end{defn}

\smallskip

As before, we interpret the collection $F$ as a mapping of $C_a$ to $C_b'$
obtained by choosing one of the morphism in the collection $\{ F_{ab,r} \}$
so that the probability of choosing $F_{ab,r}$ is $\mu^{ab}_r$.

\smallskip

\begin{rem}\label{PCzerosum}{\rm
The same argument used in Lemma~\ref{zeroobjPS} and Theorem~\ref{coproddiagr}
shows that if $\cC$ has a zero object and a categorical sum, then $\cP\cC$ also has
a zero object, given by the zero object of $\cC$ with $\Lambda=1$ and a categorical
sum given by $\Lambda C \amalg \Lambda' C' = \sum_{i,j} \lambda_i \lambda'_j C_i \amalg_\cC C'_j$,
which satisfies the the universal property, with $\Phi\amalg \Phi': \Lambda C \amalg \Lambda' C' \to \Sigma C''$
given by $\Phi\amalg \Phi' =(S \amalg S', F\amalg F')$, where $(S\amalg S')_{u,(a,a')}=\sigma_u^{-1} S_{ua} S'_{ua'}$ for $\sigma_u\neq 0$ and $(S\amalg S')_{u,(a,a')}=S_{ua}+ S'_{u,a'}$ if $\sigma_u=0$, 
and with $F\amalg F'=\{ F_{ua,r}\amalg_\cC F_{ua',r'} \}$ with probabilities 
$\sigma_u^{-1} \mu^{ua}_r \mu^{ua'}_{r'}$ for $\sigma_u\neq 0$ and $M^{-1} \mu^{ua}_r + N^{-1} \mu^{ua'}_{r'}$
for $\sigma_u=0$.
}\end{rem}

\smallskip

\begin{rem}\label{wreath}{\rm
The construction of $\cP\cC$ from $\cC$ can be seen as a wreath product 
$\cF\cP \wr \cC$ of the category $\cC$ with the category $\cF\cP$ of finite probabilities.
}\end{rem}

\medskip
\section{Information loss}

We recall here some results about information loss from \cite{BFL} that will be useful in
the following sections.

\smallskip
\subsection{Information loss}\label{InfoLoss}

In classical information, the Shannon entropy of a measure $P=(P_i)$
on a finite set of cardinality $n$ is defined as
\begin{equation}\label{ShannonS}
S(P)=- \sum_{i=1}^n P_i \log P_i .
\end{equation}
It is well known that the function \eqref{ShannonS} can
be characterized uniquely (up to an overall multiplicative
constant $C>0$) by a set of simple axioms, the
Khinchin axioms \cite{Khinchin} expressing the
properties of continuity, maximality at the
equidistribution, additivity over subsystems $S(A\cup B)=S(A) + S(B|A)$, 
and expansibility (compatibility with respect to changing the dimension $n$). 
These axioms were also formulated in a more concise way in \cite{Faddeev}.

\smallskip

The Kullback--Leibler divergence, or relative entropy of two
probability distributions $P,Q$ on the same finite set $\Sigma$ with $\# \Sigma =n$
is given by
\begin{equation}\label{KLdiv}
KL(P||Q)=- \sum_i P_i \log\frac{Q_i}{P_i}.
\end{equation}
It is not a metric (it is not symmetric, it can take infinite value, and does not
satisfy a triangle inequality) but it defines a metric (the Fisher--Rao information 
metric) when considering the leading term in the expansion $KL(P+h||P+h')$.
In the case of probability distributions $P,Q$ on different sets related through a 
map $f:\Sigma\to \Sigma'$, it is possible to compare them in a similar way via
a particular case of conditional entropy $H(P|Q)$ (see \S 3 of \cite{BFL}). 

\smallskip

More precisely, consider then the category whose objects are finite sets $\Sigma$ with
probability measures $P$ with morphisms $f:(\Sigma,P) \to (\Sigma',Q)$
given by measure preserving maps, satisfying 
\begin{equation}\label{measpres}
 Q_j = \sum_{i\in f^{-1}(j)} P_i. 
\end{equation}
Then the information loss of a morphism $f:(\Sigma,P) \to (\Sigma',Q)$ is
defined as the conditional entropy
\begin{equation}\label{infoloss}
 \cI(f)=  \sum_{s\in \Sigma} P_s \log\frac{Q_{f(s)}}{P_s} = S(P)-S(Q) . 
\end{equation} 
The last equality holds as a consequence of the measure preserving assumption \eqref{measpres}.
The setting can be generalized by considering finite measures not necessarily normalized to
probability measures, see \cite{BFL}, \cite{BF}. 
The expression in \eqref{infoloss} can be viewed as a Kullback--Leibler divergence between
$P$ and a non-normalized pullback measure of $Q$ along $f$.

\smallskip

It was proved in \cite{BFL} that 
the information loss function $\cI(f)$ of \eqref{infoloss} satisfies an axiomatic 
characterization  (up to a constant multiplicative factor), 
which follows from the
Khinchin axioms of the Shannon entropy (as reformulated in \cite{Faddeev}).
The characterizing axioms in this setting are
\begin{itemize}
\item Additivity under composition of morphisms: 
$\cI(f\circ g)=\cI(f)+\cI(g)$;
\item Additivity under direct sums: $\cI(f\oplus g)=\cI(f)+\cI(g)$;
\item Homogeneity under scaling: $\cI(\lambda f)=\lambda \cI(f)$, for $\lambda\in \R^*_+$.
\end{itemize}
The last two properties are replaced by the single additivity over convex combinations
\begin{equation}\label{convcomb}
\cI(\lambda f\oplus (1-\lambda) g)=\lambda \cI(f)+(1-\lambda) \cI(g), 
\end{equation}
for $\lambda\in [0,1]$,
if the normalization of measures is preserved, see \cite{BFL}.
Additivity under composition plays the role of a functoriality property in
the framework of \cite{BFL}.  

\medskip
\subsection{Information loss and the category of stochastic maps}

The argument of \cite{BFL} on the unique characterization of the information loss
functional can be easily adapted to the category $\cF\cP$ of finite probability measures
introduced above.

\smallskip

The sets of morphisms ${\rm Mor}_{\cF\cP}((X,P),(Y,Q))$ are convex sets, hence
in particular they are topological spaces, so we can consider continuous functions on
these sets.

\smallskip

\begin{defn}\label{infolossdef}
An information loss functional on $\cF\cP$ is a continuous real valued map on
the set of morphisms $\cH: {\rm Mor}_{\cF\cP} \to \R$ with the properties
\begin{enumerate}
\item the function $\cH(S)=0$ on isomorphisms; 
\item for all $S\in {\rm Mor}_{\cF\cP}((X,P),(Y,Q))$
and all $S'\in {\rm Mor}_{\cF\cP}((Y,Q),(Z,Q'))$
\begin{equation}\label{compose}
\cH(S'\circ S)= \cH(S') + \cH(S)
\end{equation}
\item for all $S \in {\rm Mor}_{\cF\cP}((X,P),(Y,Q))$ and $S'\in {\rm Mor}_{\cF\cP}((X',P'),(Y,Q))$
and $\lambda S\oplus (1-\lambda) S' \in {\rm Mor}_{\cF\cP}((X\sqcup X',\lambda P \oplus (1-\lambda) P'), (Y,Q))$
\begin{equation}\label{combine}
\cH(\lambda S\oplus (1-\lambda) S') = \lambda \cH(S) + (1-\lambda) \cH(S') + \cH( \hat 1_{(\lambda,1-\lambda)}) ,
\end{equation}
for $\hat 1_{(\lambda,1-\lambda)}$ the unique morphism from $(\{x,y\}, (\lambda,1-\lambda))$ to
the zero object. 
\end{enumerate}
\end{defn}

With respect to the last property listed above, note that the operation of taking a
disjoint union $X\sqcup X'$ with the weighted sum of probabilities $\lambda P \oplus (1-\lambda) P'$ 
is {\em not} the coproduct in the category. We discuss the behavior with respect to the
coproduct in Corollary~\ref{coprodIL} below. 

\smallskip

The following is essentially the same argument given in \cite{BFL}.

\begin{lem}\label{Shannon}
For $S\in {\rm Mor}_{\cF\cP}((X,P),(Y,Q))$, setting $\cH(S)=H(Q)-H(P)$,
with $H(P)=-\sum_{x\in X} P_x\log P_x$ the Shannon entropy satisfies
all the properties of Definition~\ref{infolossdef}.
\end{lem}

\proof The first three properties are clearly satisfied, since $\cH(S)$
depends only on the source and target probabilities $P,Q$, through
the difference of values of $H$ and the Shannon entropy of
finite probability distributions is invariant under isomorphisms. 
The last property is satisfied because of
the additivity over subsystems of the Shannon entropy, namely the
property that, for all probabilities $P=(P_1,\ldots, P_n)$ and $Q_i$
$$ H(P_1 Q_1 \oplus \cdots \oplus P_n Q_n) = H(P) + \sum_{i=1}^n P_i H(Q_i). $$
In particular, for $P=(\lambda,1-\lambda)$, we obtain
$$ H(\lambda P \oplus (1-\lambda) P')= H(\lambda,1-\lambda) + \lambda H(P) + (1-\lambda) H(P'). $$
This implies that, for $S\in {\rm Mor}_{\cF\cP}((X,P), (Y,Q))$ and $S'\in {\rm Mor}_{\cF\cP}((X',P'), (Y,Q))$
$$ H(\lambda S \oplus (1-\lambda) S')= H(Q) - \lambda H(P) - (1-\lambda) H(P') - H(\lambda,1-\lambda) , $$
where $-H(\lambda,1-\lambda)$ is identified with $$\cH( \hat 1_{(\lambda,1-\lambda)})= 
H(1)- H(\lambda, 1-\lambda),$$ where $H(1)=0$. 
\endproof

\smallskip

\begin{rem}\label{otheramalgs} {\rm 
Similarly, for $S \in {\rm Mor}_{\cF\cP}((X,P),(Y,Q))$ and $S'\in {\rm Mor}_{\cF\cP}((X,P'),(Y,Q))$
with $\lambda S+ (1-\lambda) S' \in {\rm Mor}_{\cF\cP}((X,\lambda P + (1-\lambda)P'), (Y,Q))$
$$ \cH(\lambda S+ (1-\lambda) S') = \lambda \cH(S) + (1-\lambda) \cH(S') + 
\cH( \hat 1_{(\lambda,1-\lambda)}) , $$
and for $S \in {\rm Mor}_{\cF\cP}((X,P),(Y,Q))$ and $S'\in {\rm Mor}_{\cF\cP}((X,P),(Y,Q'))$
and $\lambda S+ (1-\lambda) S' \in {\rm Mor}_{\cF\cP}((X,P), (Y,\lambda Q+ (1-\lambda) Q'))$ 
$$ \cH(\lambda S+ (1-\lambda) S') = \lambda \cH(S) + (1-\lambda) \cH(S') - 
\cH( \hat 1_{(\lambda,1-\lambda)}) , $$
while for $S \in {\rm Mor}_{\cF\cP}((X,P),(Y,Q))$ and $S'\in {\rm Mor}_{\cF\cP}((X,P'),(Y,Q'))$
and $\lambda S+ (1-\lambda) S' \in {\rm Mor}_{\cF\cP}((X,\lambda P + (1-\lambda) P'), (Y,\lambda Q+ (1-\lambda) Q'))$ we just have
$$ \cH(\lambda S+ (1-\lambda) S') = \lambda \cH(S) + (1-\lambda) \cH(S'). $$
}\end{rem}

\smallskip

\begin{prop}\label{infoloss}
The properties \eqref{compose} and \eqref{combine} determine uniquely the
information loss functional, up to an overall non-zero multiplicative constant $C$. 
For a morphism $S \in {\rm Mor}_{\cF\cP}((X,P),(Y,Q))$, the information loss is given by
\begin{equation}\label{infolossSh}
\cH(S) =C\cdot ( H(Q) - H(P) ),
\end{equation}
where $H$ is the Shannon entropy
$$ H(P)=- \sum_{x\in X} P_x \log P_x. $$
\end{prop}

\proof The argument is essentially the same as in \cite{BFL}. First observe that 
the composition of any morphism $S \in {\rm Mor}_{\cF\cP}((X,P),(Y,Q))$ with
the unique morphism $\hat 1_{(Y,Q)}: (Y,Q) \to (\{ x \},1)$ is 
$$ \hat 1_{(X,P)} =  \hat 1_{(Y,Q)} \circ S : (X,P) \to (\{ x \},1). $$
Thus, property \eqref{compose} gives
$$ \cH(S) = \cH(\hat 1_{(X,P)}) - \cH(\hat 1_{(Y,Q)}). $$
It then suffices to show that
$$ \tilde H(P):=-\cH(\hat 1_{(X,P)}) =C \cdot H(P) $$
is the Shannon entropy, up to a multiplicative constant.
To this purpose it is sufficient to check that $\cH(\hat 1_{(X,P)})$ satisfies
the Khinchin axioms as formulated in \cite{Faddeev}. The vanishing of $\cH(S)$ on isomorphisms $S$ implies
that $\cH(\hat 1_{(X,P)})$ is invariant under isomorphisms of $(X,P)$. 
Invariance under isomorphisms also implies that $\cH(\hat 1_{(\{ x \},1)})=\tilde H(1)=0$.
Continuity follows from the continuity of $\cH(S)$. Thus, we only
need to show that $\tilde H(P)$ satisfies the ``additivity over subsystems" property
of the Shannon entropy,
$$ \tilde H (P_1 Q_1, \ldots, P_n Q_n) = \tilde H(P) + \sum_{i=1}^n P_i \tilde H(Q_i). $$
We proceed inductively. We have
$$ \tilde H(\lambda Q, (1-\lambda) Q') = \tilde H(\lambda,1-\lambda) + \lambda \tilde H(Q)+
(1-\lambda) \tilde H(Q') $$ 
since
$$  -\tilde H(\lambda Q, (1-\lambda) Q') = \tilde H(1) - \tilde H(\lambda Q, (1-\lambda) Q')   =
- \cH( \lambda \hat 1_{(Y,Q)} \oplus (1-\lambda) \hat 1_{(Y',Q')} ), $$
with $\hat 1: (Y,Q) \to (\{x\},1)$ and $\hat 1_{(Y',Q')}: (Y',Q') \to (\{x\},1)$, and 
with $\tilde H(1)=0$. Then we write
$$ \tilde H (P_0 Q_0, P_1 Q_1, \ldots, P_n Q_n) = \tilde H (\lambda Q, (1-\lambda) P_1' Q_1,\ldots, (1-\lambda) P_n'Q_n) $$
$$ = \tilde H(\lambda,1-\lambda) + \lambda \tilde H(Q) + (1-\lambda) \tilde H(P_1' Q_1, \ldots, P_n' Q_n) $$ 
$$ = \tilde H(\lambda,1-\lambda) + \lambda \tilde H(Q) + (1-\lambda) (\tilde H(P') + \sum_i P_i' \tilde H(Q_i)). $$
We identify 
$$ \tilde H (P_0,\ldots, P_n) = \tilde H(\lambda, (1-\lambda) P') =
-\cH (\lambda \hat 1_{(\{x\},1)} \oplus (1-\lambda) \hat 1_{(X',P')}) $$ 
hence we obtain
$$ \tilde H(\lambda, 1-\lambda) + \lambda \tilde H(1) + (1-\lambda) \tilde H(P'), $$
so that 
$$ \tilde H (P_0 Q_0, P_1 Q_1, \ldots, P_n Q_n) = \tilde H(\lambda, (1-\lambda) P') + \lambda \tilde H(Q)
+ (1-\lambda) \sum_{i=1}^n P_i' \tilde H(Q_i)) $$
$$ = \tilde H(P_0,\ldots, P_n) +\sum_{i=0}^n P_i \tilde H(Q_i) .$$
\endproof

\smallskip

\begin{cor}\label{coprodIL}
The information loss functional evaluated on the coproduct $S \amalg S'$ of morphisms
$S:(X,P)\to (Y,Q)$ and $S':(X',P')\to (Y,Q)$ satisfies 
$$ \cH(S \amalg S')= \cH(S)+\cH(S') - \cH(\hat Q),  $$
with $\hat Q: (\{ x \},1) \to (Y,Q)$ the unique morphism from the zero object.
\end{cor}

\proof  We have 
$$ \cH(S \amalg S')= H(Q) - H(P) - H(P') = \cH(S)+\cH(S') - H(Q) ,  $$
where we identify $H(Q)=H(Q)-H(1)=\cH(\hat Q)$.
\endproof

\section{Gamma spaces and finite probabilities}

We have seen in the previous section that the category $\cP\cS_*$ of probabilistic
pointed sets and the category $\cF\cP$ of finite probability spaces have a zero
object and a categorical sum. Thus, one can apply to both of these categories
the Segal construction of $\Gamma$-spaces.

\smallskip
\subsection{Gamma spaces}

We recall the main idea of the Segal construction of $\Gamma$-spaces, \cite{Segal}.
A $\Gamma$-space is a functor $F: \Gamma^0 \to \Delta_*$ from the category of
pointed finite sets to the category of pointed simplicial sets.

\smallskip

Given a category $\cC$ with zero object and categorical sum, one can construct an
associated $\Gamma$-space $F_\cC: \Gamma^0 \to \Delta_*$ in the following way.
For a given pointed set $X\in \Gamma^0$ one considers the category $\Sigma_\cC(X)$
of summing functors $\Phi_X: P(X) \to \cC$, where $P(X)$ is the category with objects
the pointed subsets of $X$ and morphisms the inclusions and the functors
satisfy the summing properties
\begin{enumerate}
\item $\Phi_X (\star) =0$, the base point of $X$ maps to the zero orbject of $\cC$
\item $\Phi_X(A\cup A')=\Phi_X(A)\amalg \Phi_X(A')$ for any two points sets with
$A\cap A'=\{ \star \}$ and with $\amalg$ the categorical sum of $\cC$.
\end{enumerate}
The morphisms of $\Sigma_\cC(X)$ are natural transformations that are isomorphisms.
The simplicial set $F_\cC(X)=\cN \Sigma_\cC(X)$ is the nerve of the category of summing functors. 

\smallskip

Given a functor $F: \cC^{op}\times \cC \to \cD$, the coend
$\int^{C\in \cC} F(C,C)$ is the initial cowedge, where a cowedge to an object $X$ in $\cC$ is a
family of morphisms $h_A: A \to X$, for each $A\in \cC$, such that, for all morphisms
$f: A\to B$ in $\cC$ the following diagrams commute
$$ \xymatrix{  F(B,A) \ar[r]^{F(f,A)} \ar[d]^{F(B,f)} & F(A,A) \ar[d]^{h_A} \\
F(B,B) \ar[r]^{h_B} & X.
} $$

\smallskip

Given a $\Gamma$-space $F: \Gamma^0 \to \Delta_*$, it is possible to extend it to
an endofunctor $F: \Delta_* \to \Delta_*$. Let $X_n$ denote the finite pointed set
in $\Gamma^0$ with $\# X_n =n+1$. The endofunctor of $\Delta_*$ is obtained
\cite{BousFried} (see also \cite{Schwede}) as the functor (still denoted by $F$) that maps a pointed
simpliciat set $K \in \Delta_*$ to the coend
$$ F : K \mapsto \int^{X_n \in \Gamma^0} K_n \wedge F(X_n), $$
with natural assembly maps $K\wedge F(K')\to F(K\wedge K')$. 
Here the smash product $K_n \wedge F(X_n)$ attaches a copy of the
simplicial set $F(X_n)$ to each element in the set $K_n$, which is the
list of $n$-simplexes of $K$. Taking the coend then ensures that these are
glued together correctly according to the prescription of the face and degeneracy
maps of the simplicial set $K$.

\smallskip

The spectrum associated to the $\Gamma$-space is then obtained by considering
the simplicial sets given by the spheres $S^n=S^1\wedge \cdots \wedge S^1$,
with the simplicial structure induced by the simplicial circle $S^1=\Delta_1/\partial\Delta_1$. 
The maps $K' \wedge F(K) \to F(K'\wedge K)$ give rise to
the structure maps $S^1\wedge F(S^n) \to F(S^{n+1})$ of the spectrum defined by
the sequence of pointed simplicial sets $X_n=F(S^n)$.

\smallskip
\subsection{Cubical sets}

It is possible to reformulate homotopy constructions in terms of 
cubical sets rather than simplicial sets. Heuristically, while simplicial
sets are modeled on the combinatorics of finite sets, cubical sets
are modeled on power sets. We will see why this shift of viewpoint
is relevant to the setting of probabilistic pointed sets.

\smallskip

We recall the basic definition of cubical sets, \cite{Brown}, \cite{Jardine}, \cite{Kan}.
Let $\cI$ be the unit interval, considered as the combinatorial structure consisting of
two vertices and an edge connecting them. We also write $|\cI |=[0,1]$ for the geometric
unit interval given by its realization. Similarly, we write $\cI^n$ for the $n$-cube, and 
$|\cI^n|=[0,1]^n$ for its geometric realization, for any
$n\geq 0$, with $\cI^0$ a single point. The face maps $\delta^a_i : \cI^n \to \cI^{n+1}$,
for $a\in \{ 0,1 \}$ and $i=1,\ldots,n$ are given by
\begin{equation}\label{facecube}
\delta^a_i (t_1,\ldots, t_n)=(t_1,\ldots, t_{i-1}, a, t_i, \ldots, t_n)
\end{equation}
while the degeneracy maps $s_i : \cI^n \to \cI^{n-1}$ are given by
\begin{equation}\label{degcube}
s_i (t_1,\ldots, t_n)= (t_1,\ldots, t_{i-1},t_{i+1},\ldots,t_n). 
\end{equation}
These maps satisfy the cubical relations for $i<j$
\begin{equation}\label{cuberel1}
\delta^b_j \circ \delta^a_i = \delta^a_i \circ \delta^b_{j-1} \ \ \  \text{ and } \ \ \ 
s_i \circ s_j = s_{j-1} \circ s_i 
\end{equation}
as well as the relations
\begin{equation}\label{cuberel2}
\begin{array}{ll}
\delta^a_i \circ s_{j-1} = s_j \circ \delta^a_i & i<j \\[3mm]
s_j \circ \delta^a_i = 1 & i=j \\[3mm]
\delta^a_{i-1} \circ s_j = s_j \circ \delta^a_i & i > j 
\end{array}
\end{equation}

\smallskip

The cube category $\fC$ has objects $\cI^n$ for $n\geq 0$ and morphisms generated by
the face and degeneracy maps $\delta^a_i$ and $s_i$. 
A cubical set is a functor $C: \fC^{op} \to \cS$ to the category of sets. We write
$C_n=C(\cI^n)$. 

\smallskip

One can enlarge the category $\fC$ to a category $\fC_c$ that has additional
degeneracy maps $\gamma_i : \cI^n \to \cI^{n-1}$ called connections (see \cite{Brown}) 
\begin{equation}\label{conncube}
\gamma_i (t_1,\ldots, t_n)= (t_1,\ldots, t_{i-1}, \max\{ t_i, t_{i+1} \}, t_{i+2}, \ldots, t_n).
\end{equation}
These satisfy the relations 
\begin{equation}\label{cuberel3}
\gamma_i \gamma_j = \gamma_j \gamma_{i+1}, \, i\leq j; \ \ \  s_j \gamma_i =\left\{ \begin{array}{ll}
\gamma_i s_{j+1} & i<j \\ s_i^2=s_i s_{i+1} & i=j \\ \gamma_{i-1} s_j  & i>j \end{array}\right. ; 
\end{equation}
$$
\gamma_j \delta^a_i  = \left\{ \begin{array}{ll} \delta^a_i \gamma_{j-1}  & i<j \\ 1 & i=j,j+1,\, a=0 \\
\delta^a_j s_j  & i=j, j+1, \, a=1 \\  \delta^a_{i-1} \gamma_j & i> j+1. \end{array}\right.
$$
The connection maps $\gamma_i$ are extra degeneracies. While the usual
degeneracy maps $s_i$ identify opposite faces of a cube these additional
degeneracies identify adjacent faces. 

\smallskip

A cubical set with connection is a functor $C: \fC^{op}_c \to \cS$ to the category of sets.
 
\smallskip

The category of cubical sets has
these functors as objects and natural transformations as morphisms, that is, a
collection $\alpha=(\alpha_n)$ of morphisms $\alpha_n: C_n\to C'_n$ satisfying 
the compatibilities $\alpha \circ \delta^a_i =\delta^a_i \circ \alpha$ and
$\alpha \circ s_i =s_i \circ\alpha$ (and in the case with connection $\alpha \circ \gamma_i=
\gamma_i \circ \alpha$).

\smallskip

It is convenient to work with cubical sets with connection, as this corrects
the problem that the realization functor from cubical sets to topological 
spaces does not preserve finite products.  Moreover, when working with
cubical sets with connection, as shown in \cite{Anto},
there is a cubical classifying space and cubical nerve construction (see also
\cite{Fenn}) which is homotopy equivalent to the usual simplicial one. 

\smallskip

We refer to the cubical nerve of a category $\cC$ with the notation $\cN_{\fC} \cC$.
It is defined as the cubical set with
\begin{equation}\label{NCn}
(\cN_{\fC} \cC)_n = {\rm Fun}(\cI^n, \cC),
\end{equation}
where $\cI^n$ is the $n$-cube seen as a category with objects the
vertices and morphisms generated by the $1$-faces (edges), and
${\rm Fun}(\cI^n, \cC)$ is the set of functors from  $\cI^n$ to $\cC$.

\smallskip
\subsection{Cubical Gamma spaces}

We can consider the analog of $\Gamma$-spaces in the cubical setting.

\begin{defn}\label{cubeGamma}
Let $\Gamma=\cS_*$ be the category of pointed sets. A pointed cubical set with connection is a
functor $K: \fC^{op}_c \to \cS_*$. We denote by $K_n=K(\cI^n)$ with $\star \in K_n$ the
base point. We denote by $\Box_*$ the category of pointed cubical sets with connection,
with objects the functors $K: \fC^{op}_c \to \cS_*$ and morphisms the natural transformations.
A cubical $\Gamma$-space is a functor $F: \Gamma^0 \to \Box_*$.
\end{defn}

\smallskip

The Segal construction can be adapted to obtain cubical $\Gamma$-spaces from
categories with zero object and a categorical sum. One proceeds in the same way,
by constructing the category of summing functors $\Sigma_\cC(X)$ for finite
pointed sets $X\in \Gamma^0$, and then one takes the cubical nerve $\cN_{\fC}  \Sigma_\cC(X)$.
The resulting cubical $\Gamma$ space $F^\fC_\cC: \Gamma^0 \to \Box_*$ 
assigns to a pointed finite set $X$ the pointed cubical set with connection 
$F^\fC_\cC(X) =\cN_{\fC}  \Sigma_\cC(X)$. 

\smallskip

Since $\cN_{\fC}  \Sigma_\cC(X)$ is homotopy equivalent to the simplicial nerve
$\cN \Sigma_\cC(X)$ (see \cite{Anto}) there is no loss of generality in adopting 
this cubical setting for $\Gamma$-spaces. 

\smallskip
\subsection{The summing functors of probabilistic pointed sets} 

We consider again the category $\cP\cS_*$ of probabilistic pointed set. For a
choice of a finite pointed set $X \in \Gamma$, we construct the category of 
summing functors $\Sigma_{\cP\cS_*}(X)$. 

\begin{thm}\label{sumsPSstar}
Objects in the category $\Sigma_{\cP\cS_*}(X)$ of summing functors can
be identified with choices of a point $\Lambda=\{ \lambda_x \}_{x\in X\smallsetminus \{ \star \}}\in |\cI^N|$,
with $\# X = N+1$, with the summing functor $\Phi_\Lambda(A)=\Lambda_A X_A$ a combination
of $2^{N_A}$ pointed sets of cardinality $N_A=\# A-1$, with probability
distribution 
\begin{equation}\label{LambdaAt}
\Lambda_A =\{ (t_1,\ldots, t_{N_A})\,:\, t_x \in \{ \lambda_x, (1-\lambda_x) \} \},
\end{equation}
with morphisms given by permutations of the $2^{N_A}$ sequences 
in \eqref{LambdaAt}.
\end{thm}

\proof
Given a finite pointed set $X \in \Gamma$ with base point $\star$, let $P(X)$ be the category with
objects the pointed subsets of $X$ and morphisms given by inclusions.
The objects of $\Sigma_{\cP\cS_*}(X)$ consist of functors $\Phi: P(X) \to \cP\cS_*$
satisfying $\Phi(\star)=(\{ x\}, x)$, the zero object of  $\cP\cS_*$ with $\Lambda=1$,
and, for any sets $A,B\in P(X)$ with $A\cap B=\{ \star \}$,
\begin{equation}\label{summingPhi}
 \Phi(A\cup B) = \Phi(A) \amalg \Phi(B), 
\end{equation} 
the coproduct in $\cP\cS_*$.
In particular the condition \eqref{summingPhi} implies that, given a pointed subset $A\in P(X)$
the value $\Phi(A)$ is given by
\begin{equation}\label{PhiA}
 \Phi(A) =\amalg_{a\in A\smallsetminus \{ \star \}} \Phi(\{ \star, a \}). 
\end{equation} 
Given a morphism in $P(X)$, namely a pointed inclusion $j: A \hookrightarrow A'$,
we write $A'= A \vee B$ with $B=(A'\smallsetminus A)\cup \{ \star \}$ so that
$\Phi(A')=\Phi(A) \amalg \Phi(B)$. Then the morphism $\Phi(j)$ in $\cP\cS_*$ 
is the map $\Psi=(\cI,\cF)$ to the coproduct as in Theorem~\ref{coproddiagr} 
$$ \Phi(j)=\Psi=(\cI,\cF): \Phi(A) \to \Phi(A) \amalg \Phi(B). $$

\smallskip

In terms of probabilistic pointed sets, when we consider the union
$X_a \cup X_b$ of two sets of the form $X_a=\{ \star, a \}$ and $X_b=\{ \star, b \}$,
this means that we select for both $X_a$ and $X_b$ the point $\star$ as base
point and then we take the coproduct of pointed sets $(X_a,\star)\vee (X_b,\star)=\{\star,a,b\}$.
We can think of $X_a$ and $X_b$ as probabilistic pointed sets of the form
$\Lambda X_a = \lambda_a\, (X_a,\star) + (1-\lambda_a) (X_a,a)$
and $\Lambda X_b=\lambda_b\, (X_b,\star) + (1-\lambda_b) (X_b,b)$,  
where $\lambda_a$ and $\lambda_b$ in $[0,1]$ are the respective probabilities of
selecting $\star$ as the basepoint, instead of $a$ or $b$.  Thus, when
we describe a pointed subset $A\subset X$ as the coproduct of pointed sets
$A=\vee_{a\in A} X_a$ we are selecting for each $X_a$ the same basepoint $\star$,
when we view them as probabilistic pointed sets. Thus, $A$ can be obtained from
the probabilistic pointed sets $\{ \Lambda X_a \}_{a\in A\smallsetminus \{ \star \}}$
with probability 
\begin{equation}\label{problambdaA}
\lambda_A=\prod_{a\in A\smallsetminus \{ \star \}} \lambda_a.
\end{equation} 
If $\lambda_x\neq 0$ for all $x\in X$, 
these probabilities satisfy the multiplicative inclusion-exclusion relation
\begin{equation}\label{Ainex}
\lambda_{A\cup B} = \frac{\lambda_A \cdot \lambda_B}{\lambda_{A\cap B}} .
\end{equation}

\smallskip

Given a pointed subset $A\in P(X)$ and an assignment of values
$\{ \lambda_a \}_{a\in A\smallsetminus \{ \star \}}$, with $\lambda_a\in (0,1)$, 
there is a uniquely determined probabilistic pointed set, which we denote by 
$\Lambda_A X_A$, such that the pointed set $A$ occurs in the combination
$\Lambda_A X_A$ with probability $\prod_{a\in A\smallsetminus \{ \star \}} \lambda_a$.
For example, for $A=\{ \star, a, b\}$, the associated probabilistic pointed set is of the form
$$ \Lambda_A X_A = \lambda_a\lambda_b (\{\star,a\}\vee \{\star b\},\star\sim\star) + \lambda_a (1-\lambda_b) (\{\star,a\}\vee \{\star b\},\star\sim b) $$ $$ + (1-\lambda_a)\lambda_b (\{ \star,a\}\vee \{\star b\}, a\sim\star) +
(1-\lambda_a) (1-\lambda_b) (\{ \star,a\}\vee \{\star b\},a\sim b). $$
Similarly for $\# A =N_A+1$, the probabilistic pointed set  $\Lambda_A X_A$ is a combination of $2^{N_A}$
terms with probability distribution 
$\Lambda_A=\{ t_1\cdots t_{\# A} \,:\, t_a\in \{ \lambda_a, (1-\lambda)_a \} \}$.

\smallskip

This shows that in order to construct a summing functor $\Phi: P(X) \to \cP\cS_*$ it suffices to
assign a choice of coefficients $\{ \lambda_x \}_{x\in X\smallsetminus \{ \star \}}\in |\cI^N|$ 
with $N=\# X-1$ and that, conversely,
a summing functor determines a collection of $\lambda_x$ as the probabilities assigned to the
pointed sets $\{ \star, a\}$ in $\Phi(\{ \star, a\})=\Lambda X_a$.  Morphisms in the category
$\Sigma_{\cP\cS_*}(X)$ are natural transformations that are isomorphisms on objects, 
$\eta_A: \Phi(A)\stackrel{\simeq}{\to} \Phi'(A)$ compatible with inclusions
$j: A\hookrightarrow A'$, with $\Phi'(j)\circ \eta_A =\eta_{A'}\circ \Phi(j)$. These are
morphisms $\eta_A: \Lambda_A X_A \to \Lambda'_A X_A'$ in $\cP\cS_*$, with
$\eta_A=(S_A, F_A)$ with $S_A\in {\rm Mor}_{\cF\cP}(\Lambda_A, \Lambda'_A)$ an isomorphism
and $F_A=\{ (F_A)_{ab,r} \}$ a collection of pointed isomorphisms between the pointed 
sets in the combination $X_A$ and the pointed sets in the combination $X'_A$, with
probabilities $\sum_r \mu^{(ab)}_r=(S_A)_{ab}$. The only stochastic matrices with a stochastic
inverse are permutation matrices, hence the probabilities $\Lambda_A$ and $\Lambda'_A$ are
related by a permutation so $(S_A)_{ab}$ are either $0$ or $1$. Thus, the underlying pointed 
sets are identified by isomorphisms $F_{a,r}$ with probabilities $\sum_r \mu_r=1$. The
probability $\Lambda_A$ consists of all sequences
$t_1 \cdots t_N$ with $t_i \in \{ \lambda_{a_i}, (1-\lambda_{a_i}) \}$, for $\# A =N+1$ and
$\{ \lambda_x \}_{x\in X\smallsetminus\{ \star \}}$ specifying the summing functor $\Phi$. Thus, a permutation 
relating $\Lambda_A$ and $\Lambda_{A'}$,  
compatibly with morphisms $\Phi(j)$ and $\Phi'(j)$ induced 
by the inclusions $j:A\hookrightarrow B$, is a permutation of the $2^N$ sequences $t_1\cdots t_N$. 
\endproof

\smallskip

\begin{prop}\label{cubenerveSum}
The cubical nerve $\cN_{\fC}  \Sigma_{\cP\cS_*}(X)$ has sets 
$$K_n=(\cN_{\fC}  \Sigma_{\cP\cS_*}(X))_n={\rm Fun}(\cI^n,\Sigma_{\cP\cS_*}(X))$$ 
given by all the assignments of a pair $(\Lambda,\P_n)$
consisting of a point $\Lambda=\{ \lambda_x \}_{x\in X\smallsetminus\{ \star \}}\in |\cI^{\# X-1}|$ 
and a pointed polytope with $2^n$ vertices 
$\P_n=\{ (t_1,\ldots, t_n)\,:\, t_i\in \{ \lambda_{x_i},1-\lambda_{x_i} \}\}$,
and distinguished vertex the sequence $(\lambda_{x_1},\ldots,\lambda_{x_n})$.
\end{prop}

\proof
The cubical nerve of the category $\Sigma_{\cP\cS_*}(X)=\{ \Sigma_{\cP\cS_*}(X)_n \}$ 
is constructed by considering, for all $n\geq 0$, the set of functors $\Theta: \cI^n \to \Sigma_{\cP\cS_*}(X)$.
The objects of $\cI^n$ are the vertices of the cube $\cI^n$, namely all the sequences $v=(s_1,\ldots, s_n)$
with $s_i \in \{0,1\}$. To each vertex $v$ of $\cI^n$, a functor $\Theta: \cI^n \to \Sigma_{\cP\cS_*}(X)$ associates
an object of $\Sigma_{\cP\cS_*}(X)$, that is, a summing functor $\Phi_v: P(X) \to \cP\cS_*$, which
is specified, as above, by the choice of a point $\{ \lambda_x \}_{x\in X\smallsetminus\{ \star \}}$ 
in the realization cube $|\cI^{\# X-1}|$. 
The morphisms in $\cI^n$ are generated by the edges $e$ of the cube and the image of any of these
morphisms under $\Theta$ is a natural transformation of functors $\Phi_v$ and $\Phi_{v'}$ 
associated to adjacent vertices $\{ v, v' \}=\partial e$ of $\cI^n$. These natural transformations
are permutations, corresponding to the exchange of the sequences $v, v'\in \{0,1\}^n$,  
relating the $2^n$ sequences of $\Lambda_A X_A=\Phi_v(A)$ and 
$\Lambda'_A X'_A=\Phi_{v'}(A)$.
Thus, we can identify the datum of a functor $\Theta: \cI^n \to \Sigma_{\cP\cS_*}(X)$ with
the assignment of a point $\{ \lambda_x^v \}_{x\in X\smallsetminus\{ \star \}}\in |\cI^{\# X-1}|$ at each vertex $v\in \cI^n$ together
with a product probability space 
\begin{equation}\label{PvProd}
 \P(v)=\prod_{i=1}^n (\lambda_{x_i}^v,1-\lambda_{x_i}^v) = (\{ 0,1 \}^n, (t_1,\ldots, t_n)_{t_i\in \{ \lambda_{x_i}^v,1-\lambda_{x_i}^v\}} ). 
\end{equation} 
The set $\P(v)$ has a distinguished point given by the sequence 
$(\lambda_{x_1},\ldots,\lambda_{x_n})$, by \eqref{problambdaA}.
The sequences $\Lambda^v$ and $\Lambda^{v'}$ associated to adjacent vertices
are related by the permutation that exchanges $v=(s_1,\ldots,s_n)$ and $v'=(s_1',\ldots, s_n')$
in $\{0,1\}^n$, so that the sequences $(\lambda^v_{x_1},\ldots, \lambda^v_{x_n})$ and $(\lambda^{v'}_{x_1},\ldots, \lambda^{v'}_{x_n})$ are related by the same permutation. Adjacent vertices $v,v'$ are
sequences in $\{0,1\}^n$ that differ at a single digit. Thus, the condition implies that the corresponding
sequences $\Lambda^v$ and $\Lambda^{v'}$ also differ at a single $x_i$ where $\lambda_{x_i}$
is exchanged with $(1-\lambda_{x_i})$. We can then identify the pairs
$(\Lambda^v,\P(v))$ consisting of a point $\Lambda^v=\{ \lambda_x^v \}_{x\in X}\in |\cI^{\# X}|$ 
and a pointed polytope with $2^n$ vertices 
$\P(v)=\{ (t_1,\ldots, t_n)\,:\, t_i\in \{ \lambda_{x_i}^v,1-\lambda_{x_i}^v\}\}$, that satisfy these
permutation conditions for all adjacent vertices with a single choice of 
$\Lambda=\{ \lambda_x \}_{x\in X\smallsetminus\{ \star \}}  \in |\cI^{\# X-1}|$ (independent of $n$) and a single 
pointed polytope $\P_n=\{ (t_1,\ldots, t_n)\,:\, t_i\in \{ \lambda_{x_i},1-\lambda_{x_i} \}\}$
with base point $(\lambda_{x_1},\ldots, \lambda_{x_n})$.
\endproof

\smallskip

The geometric realization $|K|$ of $K=\{ K_n\}$ is obtained as 
$$ | K | = \bigcup_{n=1}^\infty K_n \times | \cI^n | / \sim $$
with the relation identifying faces $(x,\delta^a_i(t))\sim (\delta^a_i(x),t)$, 
degeneracies $(x, s_i(t))\sim (s_i(x),t)$, and connections 
$(x,\gamma_i(t))=(\gamma_i(x),t)$. 

\smallskip

\begin{prop}\label{geomreal}
The geometric realization $| \cN_{\fC}  \Sigma_{\cP\cS_*}(X) |$ is given by
$$ \bigcup_{Z \in |\cI^N|} |\cI_Z^N |, $$
where $\# X=N+1$ and $| \cI_Z^N |$ the geometric $N$-cube
with vertices $\{ (t_1,\ldots, t_N)\,:\, t_i\in \{ z_i, 1-z_i \}\} $, 
for $Z=\{ z_i \}_{i=1}^N\in |\cI^N|=[0,1]^N$.
\end{prop}

\smallskip

\begin{center}
\begin{figure}[h]
\includegraphics[scale=0.3]{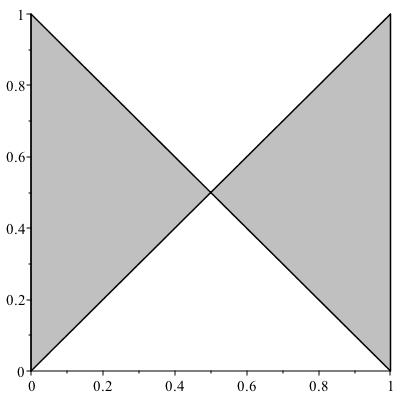} 
\caption{The realization $| \cN_{\fC}  \Sigma_{\cP\cS_*}(X) |$ for $X=\{ \star,x \}$. \label{geomrelFig}}
\end{figure}
\end{center}

\smallskip
\subsection{Smash product}

The category $\cS_*$ of pointed finite sets has a smash product operation given by
$(X,x)\wedge (Y,y)=(X\times Y/(X\times\{ y\} \cup \{ x \} \times Y), z)$ with $z$
the base point obtained from quotienting the subspace $X\times\{ y\} \cup \{ x \} \times Y$.

\smallskip

This extends to a smash product $K\wedge K'$ of pointed cubical sets $K, K':\fC_c^{op} \to \cS_*$,
as in the case of pointed simplicial sets.
 
\smallskip

The smash product can be extended to the category $\cP\cS_*$ by taking
\begin{equation}\label{smash}
\Lambda X \wedge \Lambda' X' =\sum_{ij} \lambda_i \lambda'_j (X_i\wedge Y_j, z_{ij}),
\end{equation}
for $\Lambda X=\sum_i \lambda_i (X,x_i)$ and $\Lambda'X'=\sum_j \lambda'_j (X'_j,x'_j)$.

\smallskip

Note, however, that as observed in \S \ref{prodcoprodSec}
the product of statistically independent probability spaces does not satisfy the
universal property of a categorical product.

\bigskip

\section{Stochastic Gamma spaces}

In the previous section we constructed a category $\cP\cS_*$ of probabilistic pointed sets and 
a related category $\cF\cP$ of finite probability spaces, which have 
with zero objects and a categorical sum and we have applied the Segal construction of
Gamma spaces \cite{Segal}. The resulting $\Gamma$ space $F_{\cP\cS_*}$ can be
seen as a generalization of the $\Gamma$-space $F_{\cS_*}$ associated to the
sphere spectrum, with $\cS_*$ the category of pointed finite sets.

\smallskip

Here we change the viewpoint, and we modify the Segal construction in the same
manner in which we have modified the category $\Gamma^0$ of pointed finite sets to
its probabilistic version $\cP\cS_*$. Namely we make both the notion of
Gamma spaces and the associated construction of spectra stochastic. 

\smallskip
\subsection{probabilistic pointed cubical sets}

We consider here a probabilistic version of the category $\Box_*$ of pointed cubical
sets, modeled on the probabilistic version $\cP\cS_*$ of the category $\cS_*$ of
finite pointed sets.

\smallskip

\begin{defn}\label{probBox}
The category $\cP\Box_*$ is obtained as a wreath product of the category $\Box_*$ 
and the category of finite sets $\cF\cP$ according to the general procedure of
Definition~\ref{catPC}. Namely, objects of $\cP\Box_*$ are formal convex combinations
$\Lambda K =\sum_i \lambda_i K_i$ of objects $K_i\in \cP\Box_*$, that is, functors
$K_i: \fC^{op}_c\to \cS_*$. Morphisms in $\cP\Box_*$ between objects $\Lambda K$ and
$\Lambda' K'$ are pairs $\Phi=(S,\eta)$ of a stochastic matrix $S$ with $S\Lambda=\Lambda'$
and a collection $\eta=\{ \eta_{ij,r} \}$ with $\eta_{ij,r}: K_j \to K'_i$ natural transformations
chosen with probabilities $\mu^{ij}_r$ with $\sum_r \mu^{ij}_r=S_{ij}$.
A probabilistic pointed cubical set with connection is an object in $\cP\Box_*$.
\end{defn}

\smallskip

\begin{lem}\label{probBox2}
The category $\cP\Box_*$ of probabilistic pointed cubical sets as in Definition~\ref{probBox}
can be identified with the category of functors 
\begin{equation}\label{stochBox}
\bK: \fC_c^{op} \to \cP\cS_*.
\end{equation}
with morphisms given by  natural transformations. 
\end{lem}

\proof 
An object $\Lambda K$ in $\cP\Box_*$ defines a fuctor $\bK$ with $\bK(\cI^n)=\sum_i \lambda_i K_i(\cI^n)$
in $\cP\cS_*$. Moreover, a morphism $(S,\eta=\{ \eta_{ij,r} \})$ in $\cP\Box_*$ defines a natural transformation
$\eta: \bK=\Lambda K\to \bK'=\Lambda' K'$ with $(S,\eta_n)$ the morphisms in $\cP\cS_*$ 
$\Lambda'=S\Lambda$ and $\eta_n=\{ \eta_{n,ij,r} \}: K_j(\cI^n)\to K_i(\cI^n)$.
Conversely, a functor $\bK: \fC_c^{op} \to \cP\cS_*$ given, for each $n$, an object
$\bK(\cI^n)=\Lambda^{(n)} K^{(n)} \in \cP\cS_*$. The compatibility with the faces, degeneracies, and
connections $\delta^a_i$, $\sigma_i$, and $\gamma_i$ implies that there are morphisms
$$ \Phi^\delta =(S^\delta, \underline{\delta}), \ \ \  \Phi^\sigma=(S^\sigma, \underline{\sigma}), \ \ \
\Phi^\gamma =(S^\gamma, \underline{\gamma}), $$
with $\Phi^\delta=(\Phi^{\delta^a}_i)_{u,u'}$, $\Phi^\sigma=(\Phi^\sigma_i)_{u,u'}$, and
$\Phi^\gamma=(\Phi^\gamma_i)_{u,u'}$
satisfying the cubical relations. In particular, the relations \eqref{cuberel1}, \eqref{cuberel3}
imply that the stochastic matrices $S_{\delta^a_i}$, $S_{\sigma_i}$ and $S_{\gamma_i}$
are invertible with stochastic inverses, hence all $\Lambda^{(n)} =\Lambda$ up to permutations
and all the morphisms $\underline{\delta}$, $\underline{\sigma}$, $\underline{\gamma}$
consist of a single morphism applied with probability one. Thus, up to a permutation of the terms, the
$\Phi^\delta$, $\Phi^\sigma$ and $\Phi^\gamma$ are diagonal, given by the identity 
on $\Lambda$ and the usual faces, degeneracies, and connections on the underlying
pointed cubical sets. Thus, $\bK =\Lambda K$ is an object of $\cP\Box_*$ as in Definition~\ref{probBox}
and the natural transformations are also given by morphisms in $\cP\Box_*$. 
\endproof

\smallskip

\smallskip
\subsection{Stochastic Gamma spaces}

We extend then the definition of $\Gamma$-spaces to the probabilistic setting as
follows.

\begin{defn}\label{probGamma}
A probabilistic cubical $\Gamma$-space is a functor $F: \cP\cS_* \to \cP\Box_*$.
\end{defn}

The Segal construction assigns a $\Gamma$-space $F_\cC$ to a category $\cC$
with zero objects and a categorical sum. We show here that the construction extends 
to obtain a probabilistic $\Gamma$ space. 

\smallskip
\subsubsection{Stochastic summing functors}

Given an object $\Lambda X=\sum_i \lambda_i (X_i,x_i)$ in $\cP\cS_*$, 
we consider a category $\cP(\Lambda X)$ defined as follows.

\begin{defn}\label{PLambdaX}
The category $\cP(\Lambda X)$ has objects $\Lambda A=\sum_i \lambda_i (A_i,x_i)$
with $(A_i,x_i)\in P(X_i,x_i)$, where $P(X_i,x_i)$ is the category of pointed subsets
and inclusions. Morphisms in $\cP(\Lambda X)$ are morphisms $\Phi_{A,A"}: \Lambda A\to \Lambda A'$
where the stochastic matrix $S$ is the identity and $F_{A,A'}=\{ F_{A_i,A'_i} \}$ is a collection
of pointed inclusions $F_{A_i,A'_i}: (A_i,x_i) \hookrightarrow (A'_i,x_i)$ applied with probability $1$.
\end{defn}

\smallskip

Let $\cC$ be a category with zero object and coproduct and let $\cP\cC$ be its
probabilistic category as in Definition~\ref{catPC}. 

\smallskip

\begin{defn}\label{sumfuncatP}
Let $\cC$ be a category with zero objects and a categorical sum and let
$\Lambda X$ be an object in $\cP\cS_*$. The category $\cP\Sigma_\cC(\Lambda X)$
has objects the summing functors $\Theta: \cP(\Lambda X) \to \cC$ and
morphisms the natural transformations that are isomorphisms on objects.
\end{defn}

\smallskip

\begin{defn}\label{sumPC}
A summing functor $\Theta: \cP(\Lambda X)\to \cP\cC$ is a functor satisfying
$\Theta(\sum_i \lambda_i (A_i,x_i))=\sum_i \lambda_i \Theta_i(A_i,x_i)$, where
$\Theta_i: P(X_i,x_i) \to \cP\cC$ are summing functors in $\Sigma_{\cP\cC}(X_i,x_i)$.
The category $\cP\Sigma_{\cP\cS}(\Lambda X)$ has objects the summing functors 
$\Theta: \cP(\Lambda X)\to \cP\cC$ and morphisms given by natural transformations
$\{ \eta_i \}$ of the summing functors $\Theta_i$ that are isomorphisms on objects.
\end{defn}

\smallskip

\begin{lem}\label{nervePC}
The the cubical nerve $\cN_\fC( \cP\Sigma_{\cP\cC}(\Lambda X) )$
can be identified with the probabilistic pointed cubical set 
$\sum_i \lambda_i \cN_\fC(\Sigma_{\cP\cC}(X_i,x_i))$. 
\end{lem}

\proof The cubical nerve is given by $(\cN_\fC( \cP\Sigma_{\cP\cC}(\Lambda X) )_n={\rm Fun}(\cI^n,
\cP\Sigma_{\cP\cC}(\Lambda X))$. Thus, to each vertex $v\in \cI^n$ one assigns a summing functor
$\Theta_v =\sum_i \lambda_i \Theta_{v,i}$ and to each edge $e\in \cI^n$ one assigns a natural
transformation $\eta_e=\{ \eta_{e,i} \}$. Thus, a point in $\cN_\fC( \cP\Sigma_{\cP\cC}(\Lambda X) )$
is specified uniquely by a choice of points in the $\cN_\fC(\Sigma_{\cP\cC}(X_i,x_i))$ and viceversa,
so that we can identify $\cN_\fC( \cP\Sigma_{\cP\cC}(\Lambda X) )$ with the data of a 
probabilistic pointed cubical set given by the formal combination
$\sum_i \lambda_i \cN_\fC(\Sigma_{\cP\cC}(X_i,x_i))$.
\endproof

\smallskip

\begin{lem}\label{stochGammaPC}
Given $\cC$ a category with zero object and coproduct and $\cP\cC$ its
probabilistic category as in Definition~\ref{catPC}, one obtains an associated 
probabilistic cubical $\Gamma$-space $F_{\cP\cC}: \cP\cS_* \to \cP\Box_*$ 
\end{lem}

\proof The functor $F_{\cP\cC}: \cP\cS_* \to \cP\Box_*$
assigns to an object $\Lambda X=\sum_i \lambda_i (X_i,x_i) 
\in {\rm Obj}(\cP\cS_*)$ the probabilistic pointed  cubical set 
$\sum_i \lambda_i \cN_\fC(\Sigma_{\cP\cC}(X_i,x_i))$. To a morphism 
$\Phi=(S,F)\in {\rm Mor}_{\cP\cS_*}(\Lambda X, \Lambda' X')$ with $S\Lambda=\Lambda'$
and $F=\{ F_{ij,r} \}$ with probabilities with $\sum_r \mu^{ij}_r =S_{ij}$, the functor
assigns the morphism of
probabilistic pointed cubical sets with the same stochastic matrix $S$ and with
a collection $\phi_{ij,r}: \cN_\fC(\Sigma_{\cP\cC}(X_j,x_j)) \to \cN_\fC(\Sigma_{\cP\cC}(X'_i,x'_i))$
applied with the same probabilities $\mu^{ij}_r$.
\endproof

\smallskip

\begin{rem}\label{endoFPC}{\rm
As in the case of ordinary $\Gamma$-spaces, we can extend the functors
$F: \cP\cS_*\to \cP\Box_*$ to endofunctors $F:  \cP\Box_* \to  \cP\Box_*$
by an analogous coend construction.
}\end{rem}

\smallskip
\subsection{Smash product of stochastic $\Gamma$-spaces}

In the usual setting of pointed sets and pointed simplicial sets, an
advantage of using $\Gamma$-spaces to construct spectra lies
in the fact that there is a simple construction of a smash product
for $\Gamma$-spaces, \cite{Lyd}. The main drawback of restricting
attention to spectra obtained from $\Gamma$-spaces is that
the construction only gives rise to connective spectra, \cite{Segal},
\cite{Thoma}.

\smallskip

The category $\Gamma^0$ of finite pointed sets has a smash product
functor $\land: \Gamma^0 \times \Gamma^0 \to \Gamma^0$, with 
$((X,x), (Y,y)\mapsto (X,x)\wedge (Y,y)=(X\times Y/(X\times \{ y \} \cup \{ x \}\times Y),\star)$,
which extends to a smash product $K\wedge K'$ of arbitrary pointed (simplicial) sets. 

\smallskip

The smash product of $\Gamma$-spaces constructed in \cite{Lyd} 
is obtained by first associating to a pair $F,F': \Gamma^0 \to \Delta_*$
of $\Gamma$-spaces a bi-$\Gamma$-space $F\tilde\land F': \Gamma^0\times \Gamma^0\to  \Delta_*$
$$ (F \tilde\wedge F') ((X,x),(Y,y)) = F(X,x) \wedge F'(Y,y) $$
and then defining
$$ (F\wedge F')((X,x) = {\rm colim}_{ (X_1,x_1)\wedge (X_2,x_2) \to (X,x)}
(F \tilde\wedge F') ((X_1,x_1),(X_2,x_2)), $$
where $(X_1,x_1)\wedge (X_2,x_2)$ is the smash product $\wedge: \Gamma^0\times \Gamma^0 \to \Gamma^0$. It is shown in \cite{Lyd} that, up to natural isomorphism, this smash product is 
associative and commutative and with unit given by the $\Gamma$-space $\bS$, and that
the category of $\Gamma$-spaces is symmetric monoidal with respect to this product. 

\smallskip

\begin{defn}\label{prodPS}
Given two probabilistic cubical $\Gamma$-spaces $F, F': \cP\cS_* \to \cP\Box_*$,
we set 
\begin{equation}\label{extsmashPS}
 (F \tilde\wedge F') (\Lambda X,\Lambda' Y) = F(\Lambda X) \wedge F'(\Lambda' Y) 
\end{equation}
where $\Lambda K \wedge \Lambda' K'$ in $\cP\Box_*$ is defined as in \eqref{smash} 
\begin{equation}\label{smashPBox}
\Lambda K \wedge \Lambda' K' =\sum_{i,j} \lambda_i \lambda_j K_i \wedge K'_j,
\end{equation}
with $K_i \wedge K'_j$ the smash product in $\Box_*$. One then defines
\begin{equation}\label{smashFF}
(F\wedge F')(\Lambda X) = {\rm colim}_{ \Lambda_1 X_1 \wedge \Lambda_2 X_2 \to \Lambda X}
(F \tilde\wedge F') ((\Lambda_1 X_1), (\Lambda_2 X_2)). 
\end{equation}
\end{defn}

\smallskip

The morphisms $\Lambda_1 X_1 \wedge \Lambda_2 X_2 \to \Lambda X$ are of the form
$S_{u, (a,a')}=\lambda_u^{-1} S_{ua} S'_{ua'}$ when $\lambda_u\neq 0$ and
$S_{u, (a,a')}=S_{ua} + S'_{ua'}$ otherwise and $f\wedge f' =\{ f_{ua,r} \wedge f'_{ua',r'} \}$
with probabilities $\lambda_u^{-1} \mu^{ua}_r \mu^{ua'}_{r'}$ or $M^{-1} \mu^{ua}_r + N^{-1} \mu^{ua'}_{r'}$,
respectively when $\lambda_u\neq 0$ or $\lambda_u= 0$.

\medskip
\section{Information loss on probabilistic categories}

We extend here the information loss functional on $\cF\cP$ discussed in \S \ref{InfoLoss}
to probabilistic categories $\cP\cC$ and in particular to the category $\cP\cS_*$ of probabilistic
pointed sets and the category $\cP\Box_*$ of probabilistic pointed cubical sets.

\smallskip
\subsection{Probabilistic categories and information}

Let $\cC$ be a category with zero object and coproduct and let $\cP\cC$ be the
associated probabilistic category $\cF\cP \wr \cC$ constructed as in Definition~\ref{catPC}. 

\smallskip

\begin{lem}\label{convexMorPC}
The sets of morphisms ${\rm Mor}_{\cP\cC}(\Lambda C, \Lambda' C')$ are convex sets.
\end{lem}

\proof Given $\Phi, \Phi'\in {\rm Mor}_{\cP\cC}(\Lambda C, \Lambda' C')$ and any
$\lambda \in [0,1]$ we obtain a morphism $\lambda \Phi + (1-\lambda) \Phi'$ as follows.
We have $\Phi=(S,F)$ and $\Phi'=(S',F')$ with $S, S'$ stochastic matrices with $S\Lambda =\Lambda'$
and $S'\Lambda =\Lambda'$ and $F=\{ F_{ab,r} \}$ and $F'=\{ F'_{ab,r'} \}$ with respective probabilities
$\mu^{ab}_r$ and $\nu^{ab}_{r'}$ with $\sum_r \mu^{ab}_r =S_{ab}$ and $\sum_{r'} \nu^{ab}_{r'}=S'_{ab}$.
The combination $S_\lambda=\lambda S + (1-\lambda)S'$ is a stochastic matrix with 
$S_\lambda \Lambda =\Lambda'$. We take $F_\lambda =\{ F_{ab,r} \}\cup \{ F'_{ab,r'} \}$
with probabilities $\lambda \mu^{ab}_r$ and $(1-\lambda) \nu^{ab}_{r'}$. We obtain in this
way a morphism $\lambda \Phi + (1-\lambda) \Phi'=(S_\lambda, F_\lambda) \in {\rm Mor}_{\cP\cC}(\Lambda C, \Lambda' C')$. 
\endproof

\smallskip

In the same way, we also have the following forms of convex combinations in the category $\cP\cC$.

\begin{lem}\label{otherconvexPC}
Given objects $\Lambda C$, $\Lambda' C'$ and $\Sigma C''$ in $\cP\cC$, and given $\lambda \in [0,1]$,
consider the object 
\begin{equation}\label{convexObj}
 \lambda \Lambda C + (1-\lambda) \Lambda' C' := \sum_i \lambda \lambda_i C_i 
 + \sum_j (1-\lambda) \lambda'_j C'_j . 
\end{equation} 
A choice of morphisms $\Phi\in {\rm Mor}_{\cP\cC}(\Lambda C, \Sigma C'')$ and 
$\Phi'\in {\rm Mor}_{\cP\cC}(\Lambda' C', \Sigma C'')$ determines a morphism 
\begin{equation}\label{convexmorPC}
\lambda \Phi + (1-\lambda) \Phi'
\in {\rm Mor}_{\cP\cC}(\lambda \Lambda C + (1-\lambda) \Lambda' C', \Sigma C'') 
\end{equation}
with stochastic matrix $\lambda S + (1-\lambda) S'$ and morphisms $\{ F_{ua,r} \} \cup \{ F'_{ua',r'} \}$
with probabilities $\lambda \mu^{ua}_r$ and $(1-\lambda) \mu^{ua'}_{r'}$.
Similarly, given objects $\Lambda C$, $\Sigma C'$, $\Sigma C''$, one can form an object
$\lambda \Sigma C' + (1-\lambda) \Sigma C''$ and morphisms $\lambda \Phi + (1-\lambda) \Phi' \in
 {\rm Mor}_{\cP\cC}(\Lambda C, \lambda \Sigma C' + (1-\lambda) \Sigma C'')$ with 
 $\lambda S + (1-\lambda) S'$ and with $\{ F_{ua,r} \} \cup \{ F'_{u'a,r'} \}$ with 
 probabilities $\lambda \mu^{ua}_r$ and $(1-\lambda) \mu^{u'a}_{r'}$.
\end{lem}

\smallskip

\begin{defn}\label{infolossPC}
An information loss functional 
$$ \cH: \cup_{\Lambda C, \Lambda' C'} {\rm Mor}_{\cP\cC}(\Lambda C, \Lambda' C') \to \R $$
is characterized by the properties:
\begin{enumerate}
\item vanishing on isomorphisms: $\cH(\Phi)=0$ if $\Phi$ is an isomorphism
\item additivity under composition: $\cH(\Phi\circ \Phi')=\cH(\Phi)+\cH(\Phi')$
\item extensivity under convex combinations: 
$$ \cH( \lambda \Phi + (1-\lambda) \Phi' ) = \lambda \cH(\Phi) + (1-\lambda) \cH(\Phi') + \cH(\hat 1_{(\lambda,1-\lambda)}), $$
for objects and morphisms as in \eqref{convexObj}, \eqref{convexmorPC}, 
where $\hat 1_{(\lambda,1-\lambda)}$ is the unique morphism in $\cP\cC$ from the object $\Lambda 0$, 
with $\Lambda=(\lambda, 1-\lambda)$ and $0$ the zero object of $\cC$ to the zero object 
of $\cP\cC$.
\end{enumerate}
We say that $\cH$ is a strong information loss functional if it also satisfies the property
\begin{enumerate}
\item[(4)] inclusion-exclusion on coproducts: $$\cH(\Phi \amalg_{\cP\cC} \Phi')=\cH(\Phi)+\cH(\Phi')-\cH(\widehat{\Sigma C''}),$$
for $\Phi\in {\rm Mor}(\Lambda C, \Sigma C'')$, $\Phi'\in {\rm Mor}(\Lambda' C', \Sigma C'')$ and
$\widehat{\Sigma C''}$ the unique morphism from the zero object of $\cP\cC$ to $\Sigma C''$.
\end{enumerate}
\end{defn}

\smallskip

\begin{rem}\label{embedFPPC}{\rm
Let $\cC$ be a category with zero object and categorical sum, and let $\cP\cC$ be the
associated probabilistic category. The category $\cF\cP$ of finite probabilities embeds
in $\cP\cC$ via the functor $\cJ: \cF\cP \to \cP\cC$ that maps $\Lambda=(\lambda_i) \mapsto 
\sum_i \lambda_i 0_i$, a sum of copies of the zero object of $\cC$ and morphisms $S$ to
$\Phi=(S,1)$. }\end{rem}

\smallskip

\begin{lem}\label{ShannonC}
An information loss functional on $\cP\cC$ satisfying the first three properties of Definition~\ref{infolossPC}
restricts to the embedding of $\cF\cP$ in $\cP\cC$ as $\cH(\cJ(S))=\kappa (H(\Lambda')-H(\Lambda))$,
where $H(\Lambda)=-\sum_i \lambda_i \log \lambda_i$ is the Shannon entropy, for some constant 
$\kappa \neq 0$. The induced $\cH\circ \cJ$ is a strong information loss functional.
\end{lem}

\proof This follows directly from Proposition~\ref{infoloss}, since the restriction to $\cF\cP$
satisfies the properties of an information loss functional on the category
of finite probabilities, which must then be of the form $\cH(S)=\kappa (H(\Lambda')-H(\Lambda))$,
for some $\kappa \neq 0$. The condition of inclusion-exclusion on coproducts is automatically satisfied 
by the induced information loss functional on $\cF\cP$, since for
a product of statistically independent probabilities the Shannon entropy is additive,
$H(\Lambda \Lambda')=H(\Lambda)+H(\Lambda')$, hence $\cH(S\amalg S')=
\kappa (H(\Sigma)-H(\Lambda\Lambda'))=\cH(S) +\cH(S')-\kappa H(\Sigma)$,
see Corollary~\ref{coprodIL}.
\endproof

\smallskip

\begin{lem}\label{HDeltaPC}
An information loss functional on $\cP\cC$ satisfying the first three properties of Definition~\ref{infolossPC}
must be of the form $\cH(\Phi)=\tilde H(\Lambda' C')-\tilde H(\Lambda C)$, for $\Phi\in {\rm Mor}_{\cF\cP}(\Lambda C,\Lambda' C')$, with $\tilde H: {\rm Obj}(\cP\cC) \to \R$ given by
$\tilde H(\Lambda C):= -\cH(\hat 1_{\Lambda C})$ with $\hat 1_{\Lambda C}$ the unique morphism in $\cP\cC$
between the object $\Lambda C$ and the zero object of $\cP\cC$. 
\end{lem}

\proof The argument is the same as in \cite{BFL} and in the case of $\cF\cP$ of Proposition~\ref{infoloss}, 
namely the composition $\hat 1_{\Lambda' C'}\circ \Phi =\hat 1_{\Lambda C}$ for all 
$\Phi\in {\rm Mor}_{\cP\cC} (\Lambda C,\Lambda' C')$, hence by additivity under composition
$\cH(\Phi)=\cH(\hat 1_{\Lambda C})-\cH(\hat 1_{\Lambda' C'})$.
\endproof

\smallskip

\begin{lem}\label{HextensPC}
Let $\cH$ be an information loss functional on $\cP\cC$ satisfying the first three properties of 
Definition~\ref{infolossPC}, and let $\tilde H(\Lambda C):= -\cH(\hat 1_{\Lambda C})$
be the associated functional  of Lemma~\ref{HDeltaPC}. Then $\tilde H$ satisfies the extensivity property
\begin{equation}\label{extPC}
\tilde H(\Lambda C) = \kappa H(\Lambda) + \sum_i \lambda_i \tilde H(C_i)  =\kappa ( H(\Lambda) + \sum_i \lambda_i  H(C_i) ),
\end{equation}
where $H(\Lambda)$ is the Shannon entropy and $\tilde H=\kappa H: {\rm Obj}(\cC) \to \R$ given by
$\tilde H(C)=-\cH(\hat 1_C)$ with $\hat 1_C$ the unique morphism from $C$ to the zero
object  of $\cC$. 
\end{lem}

\proof This property follows from the property of extensivity under convex combinations for
information loss functionals of Definition~\ref{infolossPC}, which combined with Lemma~\ref{ShannonC}
gives
$$ \tilde H(\lambda \Lambda C + (1-\lambda) \Lambda' C')= \lambda \tilde H(\Lambda C) +
(1-\lambda) \tilde H(\Lambda' C') + \kappa H(\lambda,1-\lambda). $$
One can then inductively show as in Proposition~\ref{infoloss} that this implies the
extensivity property \eqref{extPC}.
\endproof

\smallskip

\begin{lem}\label{HonC}
An information loss functional $\cH$ on $\cP\cC$ satisfying the first three properties of Definition~\ref{infolossPC}
induces a functional $\cH : {\rm Mor}_\cC(C,C')\to \R$ given by $\cH(f)=\tilde H(C')-\tilde H(C)$, with
$\tilde H: {\rm Obj}(\cC) \to \R$ as in Lemma~\ref{HextensPC}. This functional $\cH : {\rm Mor}_\cC(C,C')\to \R$
satisfies the properties 
\begin{enumerate}
\item vanishing on isomorphisms: $\cH(f)=0$ if $f$ is an isomorphism
\item additivity on compositions: $\cH(f\circ f')=\cH(f)+ \cH(f')$
\end{enumerate}
If $\cH$ is a strong information loss functional on $\cP\cC$ satisfying also the fourth condition of Definition~\ref{infolossPC} then $\cH : {\rm Mor}_\cC(C,C')\to \R$ also satisfies 
\begin{enumerate}
\item[(3)] inclusion-exclusion on coproducts: $\cH(f\amalg_\cC f')= \cH(f)+\cH(f') - \cH(\hat C'')$,
for $f\in {\rm Mor}_\cC(C,C'')$, $f'\in {\rm Mor}_\cC(C',C'')$, and $\hat C''$ the unique morphism in $\cC$ 
from the zero object of $\cC$ to $C''$.
\end{enumerate}
\end{lem}

\proof The properties follow directly from the properties of the information loss
functional $\cH: {\rm Mor}_{\cP\cC}(\Lambda C, \Lambda C')\to \R$ applied to
objects given by a single $C$, $C'$ in ${\rm Obj}(\cC)$ with probabilities $\Lambda=1$
and $\Lambda'=1$.
\endproof

\smallskip

In the case of categories $\cC$ that also have a product, we can
consider a more restrictive kind of information loss functionals
on $\cP\cC$, by requiring that the induced information measure
on $\cC$ of Lemma~\ref{HonC} also has a compatibility with products.

\begin{defn}\label{prodH}
If the category $\cC$ has a product $\otimes$, then a functional
$\cH(f)=\tilde H(C')-\tilde H(C)$ on the morphisms of $\cC$,
satisfying the properties of Lemma~\ref{HonC} is multiplicative if it
also has the property that $\tilde H: {\rm Obj}(\cC) \to \R$ satisfies
\begin{equation}\label{multiplH}
 \tilde H(C\otimes C')= \tilde H(C) \cdot \tilde H(C').
\end{equation}
\end{defn}

\smallskip

Note that the function $\tilde H: {\rm Obj}(\cC) \to \R$ being
multiplicative does not imply any multiplicativity property for 
$\tilde H: {\rm Obj}(\cP\cC)\to \R$, both because the product
on $\cC$ does not extend to a categorical product on $\cP\cC$,
as we previously discussed, and because the Shannon
entropy $H(\Lambda)$ behaves additively on products of
statistically independent measures, while $\tilde H(C)$
behaves multiplicatively.

\medskip
\section{Information loss and probabilistic Gamma spaces}

We show the existence of interesting information loss functionals on the
category $\cP\Box_*$ of probabilistic pointed cubical sets. By the results of
Lemma~\ref{ShannonC}, Lemma~\ref{HDeltaPC}, and Lemma~\ref{HextensPC}, 
it suffices to construct an information loss functional associated to the objects
of $\Box_*$, namely to the pointed cubical sets given by functors 
$K: \cC^{op}_c \to \cS_*$, which satisfies the three properties listed in Lemma~\ref{HonC}. 

\smallskip

The first two properties of Lemma~\ref{HonC} are satisfied by a functional of
the form $H(K')-H(K)$ (with $K$ and $K'$ the source and target of the morphism)
where $H$ is any real valued invariant of cubical (or simplicial) sets. For example,
the Euler characteristic, the Betti numbers, the chromatic number, etc. However,
the third property of Lemma~\ref{HonC}, the inclusion-exclusion relation on
coproducts restricts the possible invariants that can be used to construct
information loss functionals to invariants that behave like an Euler characteristic,
in the sense that they satisfy an inclusion-exclusion relation.

\smallskip

The Euler characteristic has an especially nice property among inclusion-exclusion
invariants, namely it is (up to a multiplicative constant) the only additive homotopy invariant
of finite CW complexes \cite{Efremov}.
Moreover, it is the only homotopy invariant that is determined
by a local formula, by adding over vertices a rational contribution given by an
alternating sum of an averaged number of $i$-cells that contain the vertex, see \cite{Levitt}.
It is unclear, however, whether a local characterization of this kind holds for invariants $\tilde H$
arising from information loss functionals.

\smallskip

\begin{lem}\label{coprodexinPS}
Let $\cH$ be a strong information loss functional on $\cP\cC$ with
$\cC=\cS_*$ the category of pointed sets. Then the inclusion-exclusion property
for coproducts implies that $\tilde H(X,\star)$ satisfies the inclusion-exclusion
relation
\begin{equation}\label{inclexclPS}
\tilde H(A\cup B,\star)=\tilde H(A,\star) + \tilde H(B,\star) - \tilde H(A\cap B,\star) 
\end{equation}
for pointed subsets $A,B$ of $(X,\star)$. It also satisfies $\tilde H(\star)=0$.
\end{lem}

\proof The relation $\cH(F\amalg F')=\cH(F)+\cH(F')-\cH(\hat Y)$ combined with
$\cH(F\amalg F')=\tilde H(Y)-\tilde H(X\amalg X')$ and $\cH(F)=\tilde H(Y)-\tilde H(X)$,
$\cH(F')=\tilde H(Y) -\tilde H(X')$ gives additivity on coproducts on objects 
\begin{equation}\label{addcoprod}
 \tilde H(X\amalg X')=\tilde H(X) + \tilde H(X'). 
\end{equation} 
We write $(A,\star)=\amalg_{a\in A\smallsetminus \{ \star \}} (\{ a,\star\}, \star)$
and $(B,\star)=\amalg_{b\in B\smallsetminus \{ \star \}} (\{ b, \star \}, \star)$ so that
$$ \tilde H(A\cup B, \star) =\sum_{x\in (A\cup B)\smallsetminus \{ \star\}} \tilde H(\{ x, \star \}, \star) $$
$$ = \sum_{a\in A\smallsetminus \{ \star \}} \tilde H(\{ a,\star\}, \star) + \sum_{b\in B\smallsetminus \{ \star \}} \tilde H(\{ b, \star \}, \star) - \sum_{x\in (A\cap B)\smallsetminus \{ \star \}} \tilde H(\{ x, \star \}, \star) $$
$$ = \tilde H(A,\star) + \tilde H(B,\star) - \tilde H(A\cap B,\star). $$
In particular $\tilde H(\star) =\tilde H((\{ \star \},\star) \amalg (\{ \star \},\star)) =\tilde H(\star) +\tilde H(\star)$,
hence $\tilde H(\star)=0$.
\endproof

\smallskip

\begin{cor}\label{solnsPS}
All the possible functionals $\tilde H(A,\star)$ satisfying \eqref{inclexclPS} are of the form
$\tilde H(A,\star) =\kappa (\# A -1)$, for some $\kappa \in \R$.
\end{cor}

\proof Consider the pointed sets of the form $(\{ x, \star\},\star)$. Let $\kappa\in \R$ be
the value $\tilde H(\{ x, \star\},\star) =\kappa$. For $\kappa=0$ one obtains $\tilde H\equiv 0$,
in which case the functional \eqref{extPC} reduces to just the Shannon entropy, while if
$\kappa\neq 0$ then 
$$ \tilde H(A,\star)=\sum_{a\in A\smallsetminus \{ \star \}} \tilde H(\{ a, \star\},\star) =
\kappa \cdot \# (A\smallsetminus \{ \star \}). $$
\endproof

\smallskip

If we drop the strong assumption for the information loss functional $\cH$ on $\cP\cS_*$
then we obtain many more possibilities besides multiples the counting of points (reduced Euler
characteristic) $\tilde\chi(X,x)=\# X-1$.

\begin{lem}\label{weakinfoPS}
Let $\cH$ be an information loss functional $\cH$ on $\cP\cS_*$ satisfying the first three properties of Definition~\ref{infolossPC}. Then the induced
$\tilde H: {\rm Obj}(\cS_*) \to \R$ as in Lemma~\ref{HextensPC} is a function $\rho(N)$ of the
cardinality $N=\#X-1$. Requiring some additional properties on the behavior on products or
coproducts in $\cS_*$ determines a more restrictive class of functions. For example:
\begin{itemize} 
\item additivity on coproducts: $\rho(N)=\kappa N$
\item multiplicativity on products: $\rho:\N \to \R^*_+$ a multiplicative semigroup
\item both of the previous properties: $\rho(N)=N$
\item additivity on products: $\rho(N)=\kappa \log(N)$ or more generally $\rho(N)=\kappa \log(\sigma(N))$
with $\sigma: \N \to \R^*_+$ a multiplicative semigroup
\item multiplicativity on coproducts: $\rho(N)=\lambda^N$.
\end{itemize}
\end{lem}

\proof The map $\tilde H: {\rm Obj}(\cS_*) \to \R$ assigns to a pointed set $(X,x)$
an invariant under isomorphisms, hence a function of the cardinality of $X$.
For the listed cases, the first is discussed Corollary~\ref{solnsPS}. The
product in $\cS_*$ is the smash product $X\wedge Y=X\times Y/(X\times \{y\}\cup \{x \}\times Y)$
with $\#(X\wedge Y)=N\cdot M$ for $N=\# X-1$ and $M=\# Y-1$, hence multiplicative
behavior on product implies $\rho(NM)=\rho(N)\rho(M)$, which means $\rho:\N \to \R^*_+$
is a multiplicative semigroup. As a multiplicative semigroup $\N$ is freely generated by the
primes, hence a semigroup homomorphism $\rho: \N \to \R^*_+$ is determined by specifying
a generator $t_p \in \R^*_+$ for each prime $p$ in $\N$. Additivity on coproducts
and multiplicativity on products imply that $\tilde H$ is the reduced Euler
characteristic $\tilde\chi(X,x)=\# X -1$.  If we require additivity on products
$\tilde H (X\wedge Y)=\tilde H(X)+\tilde H(Y)$, we have $\rho(NM)=\rho(N)+\rho(M)$, which
is satisfied by functions of the form $\rho(N)=\kappa \log(\sigma(N))$ with $\sigma: \N \to \R^*_+$ 
a multiplicative semigroup. Multiplicative behavior on coproducts is satisfied by exponentiation of
any invariant that is additive on coproducts hence by functions of the form $\rho(N)=\lambda^N$.
\endproof

\smallskip

\begin{lem}\label{coprodPBox}
Let $\cH$ be a strong information loss functional on $\cP\cC$ with
$\cC=\Box_*$ the category of pointed cubical sets with connection.
The inclusion-exclusion property for coproducts implies that $\tilde H(K)$ 
satisfies the inclusion-exclusion relation
\begin{equation}\label{tildeHie}
 \tilde H(K\cup K') = \tilde H(K) + \tilde H(K') -\tilde H(K\cap K'), 
\end{equation} 
for $K,K'$ such that $K\cap K'$ is also a pointed cubical sets with connection.
\end{lem}

\proof The inclusion-exclusion property
for coproducts on morphisms implies, as above, additivity for coproduct of objects,
$$ \tilde H(K \amalg_{\Box_*} K')= \tilde H(K) + \tilde H(K'), $$
where $K \amalg_{\Box_*} K'= K\vee K'$. Then using the previous lemma applied
to $(K\cup K')(\cI^n)=K_n \cup K'_n \in \cS_*$ we obtain the stated inclusion-exclusion 
relation.
\endproof

\smallskip

The fact that the invariants $\tilde H$ that arise from information loss functionals
satisfy an inclusion-exclusion relation is reasonable in the context of information
theory. Indeed, mutual information satisfies an inclusion-exclusion relation, hence
we can regard the expression \eqref{extPC}, where $\tilde H$ satisfies an inclusion-exclusion
relation represented by the additivity \eqref{addcoprod} with respect to the coproduct in $\cC$,
as a generalization of the inclusion-exclusion property of a mutual information measure. 

\smallskip

\begin{rem}\label{htpyinv}{\rm
Under the hypotheses of Lemma~\ref{coprodPBox}, if we also know that
$\tilde H(K)$ is a homotopy invariant, then the additivity on coproducts 
$\tilde H(K\vee K')=\tilde H(K)+\tilde H(K')$
(hence the inclusion-exclusion \eqref{tildeHie}) together with the property that 
$\tilde H(\{ x, \star \},\star)=\kappa$ imply that $\tilde H(K)=\kappa\cdot \tilde\chi(K)$ 
where $\tilde\chi(K)=\chi(K)-1$ is the reduced Euler characteristic. In particular, if
$\kappa=1$, this implies that $\tilde H(K)=\tilde\chi(K)$ is also multiplicative
under smash products $\tilde\chi (K\wedge K')=\tilde\chi(K)\tilde\chi(K')$, as in
Definition~\ref{prodH}. 
}\end{rem}

\smallskip

The characterization above of the reduced Euler characteristic can also
be restated as the characterization as the unique $\Z$ valued function
on finite pointed CW complexes that satisfies $\tilde\chi (\{ x, \star\},\star)=1$ and
$\tilde\chi(K)=\tilde\chi(A)+\tilde\chi(B)$ for any cofiber sequence $A\to K \to B$,
see Theorem~28.85 of \cite{Strom}.

\medskip
\subsection{Information loss functionals and probabilistic Gamma spaces}

Here we consider a fixed information loss functional $\cH: \cP\Box_* \to \R$ 
with sufficiently good properties and we obtain other information loss functionals
on $\cP\cS_*$ and $\cP\Box_*$ (with weaker properties) obtained by precomposing
with probabilistic $\Gamma$ spaces $F_{\cP\cC}: \cP\cS_* \to \cP\Box_*$
associated to probabilistic categories $\cP\cC$. 

\smallskip

\begin{lem}\label{problossF}
Let $\cH: \cP\Box_* \to \R$ be an information loss functional satisfying
the first three properties of Definition~\ref{infolossPC}. Let $F_{\cP\cC}: \cP\cS_*\to \cP\Box_*$
be a probabilistic $\Gamma$-space obtained from a probabilistic category $\cP\cC$ as in Lemma~\ref{stochGammaPC}, and let $\hat F_{\cP\cC}: \cP\Box_*\to \cP\Box_*$ be its
extension to an endofunctor of $\cP\Box_*$. Then the compositions $\cH\circ F_{\cP\cC}$
and $\cH\circ \hat F_{\cP\cC}$ are also information loss functionals, respectively on 
$\cP\cS_*$ and $\cP\Box_*$,
satisfying the first three properties of Definition~\ref{infolossPC}. 
\end{lem}

\proof By functoriality $F_{\cP\cC}$ maps compositions of morphisms to compositions, hence
the additivity of $\cH$ under composition is preserved by precomposing with $F_{\cP\cC}$. 
Vanishing on isomorphisms is also preserved for the same reason. Extensivity under
convex combination is preserved because, as seen in Lemma~\ref{nervePC},
the functor $F_{\cP\cC}$ assigns to an object $\Lambda X=\sum_i \lambda_i (X_i,x_i)$ in 
$\cP\cS_*$ the object in $\cP\Box_*$ given by the combination $\sum_i \lambda_i K_i$
with $K_i =\cN_\fC(\Sigma_{\cP\cC}(X_i,x_i))$. Thus, in particular it satisfies
$$F_{\cP\cC}(\lambda \Lambda X + (1-\lambda) \Lambda' X')=\lambda F_{\cP\cC}(\Lambda X)+(1-\lambda)
F_{\cP\cC}(\Lambda' X'),$$ and the extensivity of $\cH$ then gives
$$\cH(\lambda F_{\cP\cC}(\Lambda X)+(1-\lambda)
F_{\cP\cC}(\Lambda' X')) = \lambda \cH(F_{\cP\cC}(\Lambda X))+ (1-\lambda) \cH(F_{\cP\cC}(\Lambda' X'))
+\cH(\hat 1_{(\lambda,1-\lambda)}).$$
\endproof

\smallskip

\begin{rem}\label{noincex}{\rm
The compositions $\cH\circ F_{\cP\cC}$
and $\cH\circ \hat F_{\cP\cC}$ in general do not satisfy the strong condition of Definition~\ref{infolossPC},
even if $\cH$ is a strong information loss functional on $\cP\Box_*$. This can be seen in the
case of $\cC=\cS_*$. The description of the pointed cubical sets
$\cN_\fC(\Sigma_{\cP\cS_*}(X,x))$ in Proposition~\ref{cubenerveSum} 
and Proposition~\ref{geomreal} shows 
that $\cN_\fC(\Sigma_{\cP\cS_*}((X,x)\vee (Y,y))$ is not additive in $(X,x)$ and $(Y,y)$
hence even if $\cH$ satisfies the strong condition of Definition~\ref{infolossPC} the
composition $\cH \circ F_{\cP\cS_*}$ does not. 
}\end{rem}

\smallskip

 Let $\cH: \cP\Box_* \to \R$ be an information loss functional satisfying the
 first three properties of Definition~\ref{infolossPC}, given by a difference
 of invariants of target and source objects of the form
 \begin{equation}\label{logchi}
 \tilde H(\Lambda K)= H(\Lambda) + \sum_i \lambda_i H(K_i), \ \ \ \text{ with } \ \ \ H(K) = \log \tilde\chi(K),
 \end{equation}
 with $H(\Lambda)$ the Shannon entropy and 
 with $\tilde\chi(K)$ the reduced Euler characteristic. Clearly this
 information loss functional does not satisfy the inclusion-exclusion
 property of the strong condition. It satisfied instead an additivity
 property on products
 \begin{equation}\label{addprod}
 H(K\wedge K') = H(K) + H(K'),
 \end{equation}
 which follows from the multiplicative property of the reduced Euler characteristic.
 It also satisfies homotopy invariance, since it factors through the reduced Euler
 characteristic. 
 
 \smallskip
  
 Consider an information loss functional as above, with \eqref{logchi}, so that it satisfies the
 additivity property \eqref{addprod}. Consider probabilistic $\Gamma$-spaces 
 $F: \cP\cS_* \to \cP\Box_*$ of the form
 \begin{equation}\label{FsmashK}
  F(\Lambda X) = \Lambda' K' \wedge F_{\cP\cS_*}(\Lambda X),
\end{equation} 
for $\Lambda' K'$ a given stochastic pointed cubical set in $\cP\Box_*$.  These
generalize in our probabilistic setting the classical $\Gamma$-spaces $F: \Gamma^0 \to \Delta_*$
of the form $F(X)= K\wedge F_{\Gamma^0}(X)$, with $F_{\Gamma^0} : \Gamma^0\hookrightarrow \Delta_*$,
whose associated spectrum, obtained via the Segal construction, is the
suspension spectrum of the simplicial set $K$. 
These probabilistic $\Gamma$-spaces represent a product of two statistically independent systems,
$\Lambda' K'$ and $F_{\cP\cS_*}(\Lambda X)$ and the chosen information measure accordingly
splits additively
$$ \tilde H(F(\Lambda X) )=\tilde H(\Lambda' K') + \tilde H(F_{\cP\cS_*}(\Lambda X)). $$
Conversely, any probabilistic $\Gamma$-space $F: \cP\cS_* \to \cP\Box_*$ with the
property that $H(F(\Lambda X))=\alpha + H(F_{\cP\cS_*}(\Lambda X))$ for some fixed
$\alpha\in \R$ independent of $X$ and $\Lambda$ and for all $\Lambda X\in \cP\cS_*$
should be regarded from the information point of view as equivalent to a product of
two statistically independent systems, one of which is the basic $F_{\cP\cS_*}(\Lambda X)$.

 \medskip
 \section{Quantum Information and Categories}
 
 The probabilistic category $\cP\cC$ associated to a category $\cC$ with zero object and sum
 can be seen as a wreath product of the category $\cC$ and the category $\cF\cP$ of finite
 classical probabilities, hence as a probabilistic version of the category $\cC$, in the context
 of classical probability. We consider here a similar approach that associates to a category 
 $\cC$ with zero object and sum a probabilistic category $\cQ\cC$ based on quantum
 rather than classical probability. 
 
 \smallskip
 \subsection{The category of quantum probabilities}
 
We assign to a finite set $X$ a Hilbert space $\cH_X=\oplus_{x\in X} \C_x$ 
with $\C_x$ a one-dimensional space at the site $x\in X$. More generally,
we can replace the $\C_x$ with copies of a fixed finite dimensional
Hilbert space $\cV$ of a fixed dimension, which represents the internal 
degrees of freedom at the site $x\in X$.

\smallskip

\begin{defn}\label{catQF}
The category $\cF\cQ$ of finite quantum probabilities
has objects given by pairs pairs $(X,\rho_X)$ of a finite set $X$
and a density matrix $\rho_X$ on the finite dimensional Hilbert space $\cH_X$,
that is, a linear operator on $\cH_X$ satisfying $\rho_X^*=\rho_X$, 
$\rho_X\geq 0$, and $\Tr(\rho_X)=1$. The morphisms 
${\rm Mor}_{\cF\cQ}((X,\rho_X),(Y,\rho_Y))$ are given by 
quantum channels $\Phi$, that is, completely positive trace preserving maps
with $\Phi(\rho_X)=\rho_Y$.
\end{defn}

\smallskip

Quantum channels $\Phi$ can always be written (non-uniquely) in Kraus form as
$$ \Phi(\rho) =\sum_i A_i \rho A_i^*, \ \ \  \text{ with } \ \  \sum_i A_i^* A_i =1. $$
One can also represent completely positive trace preserving maps with $\Phi(\rho_X)=\rho_Y$
through the associated stochastic Choi matrix $S_\Phi$ with
\begin{equation}\label{Choi}
(\rho_Y)_{ij} =\sum_{a,b} (S_\Phi)_{\substack{ab \\ ij}}\, (\rho_X)_{ab}
\end{equation}
Kraus representations can be obtained from factorizations $S_\Phi=A A^*$.

\smallskip
\subsection{Quantum probabilistic categories $\cQ\cC$}

As in the case of classical probability, given a category $\cC$ with zero object and
sum, we construct a probabilistic version $\cQ\cC$ that maintains the same properties, but
in this case based on the quantum probabilities of $\cF\cQ$ rather than on the
classical probabilities of $\cF\cP$. 

\begin{defn}\label{QCcat}
The category $\cQ\cC$ has objects 
given by $\rho C=((C_a,C_b), \rho_{ab})_{ab}$, where $(C_a,C_b)$ is a finite collection
of pairs of objects in $\cC$ indexed over a finite set $a,b=1,\ldots,N$ and $\rho=(\rho_{ab})$
is an $N\times N$ density matrix.  
For $\rho C=((C_a,C_b), \rho_{ab})$ and $\rho' C'=((C'_i,C'_j), \rho'_{ij})$, the morphisms $\Xi \in {\rm Mor}_{\cQ\cC}(\rho C, \rho' C')$
are given by a finite collection
$$ \Xi= \{ (\phi_{ai,r},\psi_{bj,r}) \}, (S_{\Phi_r})_{\substack{ab \\ ij}}  \} $$
where $\sum_r S_{\Phi_r}=S_\Phi$ is the Choi matrix of a quantum channel $\Phi$ with $\Phi(\rho)=\rho'$.
The composition of morphisms $\Xi'\circ \Xi$ is given by the collection
$$ \Xi'\circ \Xi =\{ (\phi_{ua,r'}\circ \phi_{ai,r}, \psi_{vb,r'}\circ  \psi_{bj,r}) , (S_{\Phi_r})_{\substack{ab \\ ij}} (S_{\Phi'_{r'}})_{\substack{ij \\ uv}} \} $$
which satisfies $\sum_{r,r',i,j} (S_{\Phi_r})_{\substack{ab \\ ij}} (S_{\Phi_{r'}})_{\substack{ij \\ uv}}
=\sum_{i,j} (S_\Phi)_{\substack{ab \\ ij}} (S_{\Phi'})_{\substack{ij \\ uv}} = 
(S_{\Phi'\circ\Phi})_{\substack{ab \\ uv}}$.
\end{defn}

\smallskip

\begin{rem}\label{QCnotation}{\rm
The notation for the objects of $\cQ\cC$ in the form
$\rho C=((C_a,C_b), \rho_{ab})_{ab}$, for $a,b=1,\ldots,N$, 
includes the case where $N=1$. In this case the objects 
are just single objects $C\in {\rm Obj}(\cC)$ with weight $\rho=1$
and morphisms in $\cQ\cC$ between two objects of this form 
are morphisms in $\cC$. This embeds the category $\cC$ into
its quantum probability version $\cQ\cC$, as in the case
of the classical probabilities.
}\end{rem}

\smallskip

\begin{rem}\label{coherence}{\rm
As usual in quantum information, one interprets the off-diagonal terms $\rho_{ij}$
of a density matrix $\rho$ as describing the interference between the
amplitudes of the $i$-th and $j$-th state, hence a measure of coherence of the mixed state.
Thus, the objects $\rho C$ of the category $\cQ\cC$ have an assigned amount of coherence 
of pairs of objects in $\cC$, described by the coefficients $\rho_{ij}$ of a density matrix. The morphisms
in $\cQ\cC$ also correspond to pairs of morphisms in $\cC$ with assigned coherence,
but also transform the density matrix of the source to that of the target through a quantum channel
obtained as the combined coherence measures of all the pairs in the collection.
}\end{rem}

\smallskip

\begin{prop}\label{QCcatzerosum}
Let $\cC$ be a category with zero object and sum. Then the category $\cQ\cC$
of Definition~\ref{QCcat} also has a zero object and categorical sum. The zero
object is given by the pair $(0,1)$ with $0$ the zero object of $\cC$ with $\rho=1$ 
and the coproduct is of the form
\begin{equation}\label{sumQC}
\rho C \amalg \rho' C' = (C_i \amalg_\cC C'_j, \rho\otimes\rho') .
\end{equation}
\end{prop}

\proof The argument is analogous to the case of classical probabilities that we discussed
previously. The zero object of $\cQ\cC$ is given by the pair $(0,1)$ with $0$ the
zero object of $\cC$ and $\rho=1$. There is a unique morphism in $\cQ\cC$
from $(0,1)$ to an object $\rho C=((C_i,C_j),\rho_{ij})$ given by the unique morphisms 
$0 \to C_i$ in $\cC$ and $\Phi_{\substack{ij \\ 0}}=\rho_{ij}$.
The unique morphism from an object $\rho C=((C_i,C_j),\rho_{ij})$ to the zero
object similarly consists of the unique morphisms $C_i \to 0$ in $\cC$ and
$\Phi_{\substack{0\\ ij}}=\delta_{ij}$, which gives $\sum_{ij}\Phi_{\substack{0\\ ij}}\rho_{ij}=\Tr(\rho)=1$.
The universal property of the coproduct is satisfied with maps
$$ \xymatrix{
& ((C_u, C_s), \tilde\rho_{us}) & \\
((C_i,C_j), \rho_{ij}) \ar[ur]^{((\phi_{ri},\psi_{sj}), \Phi_1)} \ar[r]_{((\cI_i, \cI_j),\Psi)} & 
(C_i \amalg_\cC C'_j, \rho\otimes\rho') \ar[u] & 
((C_a,C_b), \rho'_{ab}) \ar[ul]_{((\phi_{ua},\psi_{sb}), \Phi_2)} \ar[l]^{(\cI_a, \cI_b),\Psi')} 
} $$
where $\cI_i : C_i \to C_i \amalg_\cC C'_j$ are the maps of the universal property
of the coproduct in $\cC$ and 
the maps $\Psi \rho = \rho\otimes \rho'$ and $\Psi' \rho' = \rho \otimes \rho'$ are given by
$$ \Psi_{\substack{ij \\ (i'j'),(ab)}}=\delta_{ii'} \delta_{jj'} \rho'_{ab} \ \ \ \text{ and } \ \ \ 
\Psi'_{\substack{ab \\ (ij), (a'b')}}=\delta_{aa'} \delta_{bb'} \rho_{ij}$$
The map $\rho C \amalg \rho' C' \to \tilde \rho \tilde C$ that makes the diagram commute 
is then given, at the level of the quantum channels, by 
\begin{equation}\label{coprodPhi1}
 \tilde\Phi_{\substack{us \\ (ij),(ab)}}  
= \tilde\rho_{us}^{-1} (\Phi_1)_{\substack{us \\ ij}} (\Phi_2)_{\substack{us \\ ab}} 
\end{equation}
when the entry $\tilde\rho_{us}\neq 0$ and 
\begin{equation}\label{coprodPhi2}
\tilde\Phi_{\substack{us \\ (ij),(ab)}}   = (\Phi_1)_{\substack{us \\ ij}} \, \delta_{ab} +  (\Phi_2)_{\substack{us \\ ab}} \, \delta_{ij} 
\end{equation}
when the matrix entry $\tilde\rho_{us} = 0$. Indeed this gives for $\tilde\rho_{us}\neq 0$
$$ \sum_{(i',j'),(a,b)}  \tilde\Phi_{\substack{us \\ (i'j'),(ab)}}  \Psi_{\substack{(i'j'),(ab) \\ (ij)}} = 
 (\sum_{i',j'} (\Phi_1)_{\substack{us \\ i'j'}} \delta_{ii'} \delta_{jj'} )\cdot 
 \tilde\rho_{us}^{-1} (\sum_{a,b} (\Phi_2)_{\substack{us \\ ab}} \rho'_{ab} ) = (\Phi_1)_{\substack{us \\ ij}} $$
and for  $\tilde\rho_{us} = 0$ it gives
$$  \sum_{i',j',a,b} ( (\Phi_1)_{\substack{us \\ i'j'}} \delta_{ab} \delta_{ii'} \delta_{jj'} \rho'_{ab} +
(\Phi_2)_{\substack{us \\ ab}} \delta_{ij} \delta_{ii'} \delta_{jj'} \rho'_{ab} ) $$
$$ = (\sum_a \rho'_{aa}) (\Phi_1)_{\substack{us \\ ij}} + 
\sum_{ab} (\Phi_2)_{\substack{us \\ ab}}\rho'_{ab} \delta_{ij} $$ 
which is just equal to $(\Phi_1)_{\substack{us \\ ij}}$, because $\Tr(\rho')=1$ and
$\sum_{ab} (\Phi_2)_{\substack{us \\ ab}}\rho'_{ab}= \tilde\rho_{us} = 0$.
The case of composition with $\Psi'$ is analogous.  
At the level of the morphisms, one considers the  
coproducts $(\phi_{ui,r} \amalg_{\cC} \phi_{ua,r'}, \psi_{sj,r}\amalg_\cC \psi_{sb,r'})$,
with $r=1,\ldots,N$ and $r'=1,\ldots,M$, in $\cC$ weighted with 
$$ \tilde\rho_{us}^{-1} (\Phi_{1,r})_{\substack{us \\ ij}} (\Phi_{2,r'})_{\substack{us \\ ab}}
, \ \ \ \text{ for } \ \ \ \tilde\rho_{us}\neq 0 $$
$$ (\Phi_{1,r})_{\substack{us \\ ij}} \, \frac{\delta_{ab}}{M} +  (\Phi_{2,r'})_{\substack{us \\ ab}} \, \frac{\delta_{ij}}{N} , \ \ \ \text{ for } \ \ \ \tilde\rho_{us}=0. $$
The rest of the argument is analogous to Theorem~\ref{coproddiagr}. 
\endproof

\smallskip

\begin{rem}\label{prodcoprodFQ}{\rm 
Note that, as in the case of classical probabilities, the coproduct induced by \eqref{sumQC}
on the category $\cF\cQ$ of finite quantum probabilities is just the product of independent
systems $\rho \amalg_{\cF\cQ} \rho' = \rho \otimes \rho'$. }
\end{rem}

\smallskip

We can identify a decoherence subcategory of $\cQ\cC$ that corresponds to the case
of mixed states with diagonal density matrices (in a fixed basis). This can be 
described the following category. 

\begin{defn}\label{decoQC}
The decoherence subcategory $\P\cC$ has objects given by pairs $(C,z)=((C_1,\ldots, C_n), (z_1: \cdots :z_n))$ with $C_i\in {\rm Obj}(\cC)$ and $z=(z_1: \cdots :z_n)\in \P^{n-1}(\C)$ with morphisms given by
a morphism $\Phi: \P^{n-1} \to \P^{m-1}$ induced by a linear map $\tilde\Phi:\C^n\to \C^m$ up to scalars
with $\Phi z = z'$ and a collection $\{ (\tilde\phi_{ji,r}: C_i \to C'_j , \tilde\Phi_r )\}$ with 
$\sum_r \tilde\Phi_r=\tilde \Phi$. 
The coproduct is given by $(C,z)\amalg (C',z')=((C_i\amalg C_j)_{ij}, \alpha_{n,m}(z,z'))$ where
$\alpha_{n,m}: \P^{n-1} \times \P^{m-1} \to \P^{nm-1}$ is the Segre embedding. 
\end{defn}

\smallskip

In particular, in this case one can interpret the objects $(C,z)$ as a superposition of the objects $C_i$
where the probability of observing $C_i$ is $|z_i|^2$.

\medskip
\subsection{A variant: categories of arrows}

A variant on the construction of the categories $\cQ\cC$ considered in the previous section
can be obtained by working with arrows of $\cC$ instead of pairs of objects in $\cC$. We
illustrate this version of the construction here. The results in the following sections
apply to both the categories $\cQ\cC$ constructed above and the categories $\cQ\cA\cC$
constructed here. 

\smallskip

First we associate to a category $\cC$ with zero object and sum another category
$\cA\cC$ with objects the morphisms of $\cC$.

\begin{defn}\label{ACcat}
The category $\cA\cC$ has objects $\phi_{C,C'}$ given by elements of ${\rm Mor}_{\cC}(C,C')$
for arbitrary $C,C'\in {\rm Obj}(\cC)$ and morphisms $L\in {\rm Mor}_{\cA\cC}(\phi_{C,C'}, \phi_{A,A'})$ 
given by pairs $L=(L_1,L_2)$ with $L_1 \in {\rm Mor}_{\cC}(C,A)$ and $L_2\in {\rm Mor}_{\cC}(C',A')$
such that the diagram commutes
$$ \xymatrix{  C \ar[r]^{\phi_{C,C'}} \ar[d]_{L_1} & C' \ar[d]^{L_2} \\
A \ar[r]_{\phi_{A,A'}} & A'
} $$
\end{defn}

\smallskip

\begin{lem}\label{zerosumAC}
If the category $\cC$ has zero object and categorical sum then the category $\cA\cC$ also does.
The zero object of $\cA\cC$ is the identity morphism $1_0$ of the zero object of $\cC$ and the
coproduct $\phi_{C,C'} \amalg_{\cA\cC} 
\phi_{A,A'}$ is given by the unique morphism $\phi_{C\amalg A, C'\amalg A'}:
C\amalg_\cC A \to C'\amalg_\cC A'$ determined by the morphisms $\phi_{C,C'}$ and $\phi_{A,A'}$.
\end{lem}

\proof There is a unique morphism $L=(L_1,L_2)$ from any $\phi_{C,C'}$ to the zero object $1_0$
with $L_1$ the unique morphism in $\cC$ from $C$ to the zero object and $L_2$ the 
unique morphism in $\cC$ from $C'$ to the zero object. Similarly, there is a unique morphism
from the zero object $1_0$ to any $\phi_{C,C'}$ with $L_1$ the unique morphism in $\cC$
from the zero object to $C$ and $L_2$ the unique morphism in $\cC$
from the zero object to $C'$, hence $1_0$ is a zero object in $\cA\cC$. Consider the
morphisms $L_C: C \to C\amalg_{\cC} A$ and $L_A: A \to C\amalg_\cC A$ in $\cC$ 
that satisfy the universal property of the coproduct in $\cC$. Similarly, consider 
$L_{C'}: C' \to C'\amalg_{\cC} A'$  and $L_{A'}: A' \to C' \amalg_{\cC} A'$. Given morphisms
$\phi_{C,C'}$ and $\phi_{A,A'}$, by the universal property of the coproduct in $\cC$
there is a unique morphism from $C\amalg_{\cC} A$ to $C' \amalg_\cC A'$ such that
the diagram commutes
$$ \xymatrix{ & A\amalg_\cC A' & \\
C \ar[ur]^{L_{C'}\circ \phi_{C,C'}} \ar[r]_{L_C} & C\amalg_{\cC} A \ar[u] &
A \ar[ul]_{L_{A'}\circ \phi_{A,A'}} \ar[l]^{L_A} 
}$$
We show that the morphism  $\phi_{C,C'}\amalg_{\cA\cC} \phi_{A,A'}: C\amalg_\cC A \to C'\amalg_\cC A'$
obtained in this way is the coproduct in $\cA\cC$ by showing that is satisfies the universal property.
Suppose given morphisms in $\cA\cC$ 
$$ \xymatrix{  C \ar[r]^{\phi_{C,C'}} \ar[d]_{L_1} & C' \ar[d]^{L_1'} \\
R \ar[r]^{\psi} & R'
}  \ \ \ \text{ and } \ \ \ 
\xymatrix{
A \ar[r]^{\phi_{A,A'}} \ar[d]_{L_2} & A' \ar[d]^{L_2'} \\
R \ar[r]^{\psi} & R'
}  $$
By the universal property of the coproduct in $\cC$ there are unique morphisms
$L: C\amalg_\cC A \to R$ and $L': C'\amalg_{\cC} A' \to R'$ such that $L\circ L_C=L_1$
and $L\circ L_A=L_2$ and $L'\circ L_{C'}=L_1'$ and $L'\circ L_{A'}=L_2'$. The diagram
$$ \xymatrix{ C\amalg_\cC A \ar[rr]^{\phi_{C,C'}\amalg_{\cA\cC} \phi_{A,A'}}  \ar[d]_{L} & &
C'\amalg_\cC A' \ar[d]^{L'} \\
R \ar[rr]^{\psi} & & R' } $$
commutes because both $\psi\circ L$ and $L\circ (\phi_{C,C'}\amalg_{\cA\cC} \phi_{A,A'})$
have the property that they give a vertical arrow that makes the following diagram commutative
$$ \xymatrix{   &  R'  & \\
C \ar[r]_{L_C} \ar[ur]^{\psi\circ L_1}       & C\amalg_\cC A \ar[u] &   A \ar[l]^{L_A} \ar[ul]_{\psi\circ L_2}
} $$
as one can see by replacing $\psi \circ L_1= L_1' \circ \phi_{C,C'}$ and $\psi \circ L_2= L_2' \circ \phi_{A,A'}$.
By the universal property of the coproduct in $\cC$ there is a unique morphism
with this property, hence $\psi\circ L=L\circ (\phi_{C,C'}\amalg_{\cA\cC} \phi_{A,A'})$.
Thus, $(L,L'): \phi_{C,C'}\amalg_{\cA\cC} \phi_{A,A'}\to \psi$ constructed in this way is a 
morphism in $\cA\cC$, and it is the unique morphism such that $(L,L')\circ (L_C,L_A)=(L_1,L_2)$ and
$(L,L')\circ (L_{C'},L_{A'})=(L_1',L_2')$. This shows that the coproduct in $\cA\cC$ satisfies the
universal property. 
\endproof

\smallskip

Then we associate to the category of arrows $\cA\cC$ a category $\cQ\cA\cC$,
which can be seen as a wreath product of $\cA\cC$ and the category of finite
quantum probabilities $\cF\cQ$, defined as follows.

\smallskip

\begin{defn}\label{QACcat}
The category $\cQ\cA\cC$ has objects $\rho \phi=\{ \phi_{ij}, \rho_{ij} \}$ given by collections
of morphisms $\phi_{ij}: C_i \to C_j$ in $\cC$, for $i,j=1,\ldots, N$ for any $N\in \N$, together
with an $N\times N$ density matrix $\rho=(\rho_{ij})$. Morphisms 
${\rm Mor}_{\cQ\cA\cC}(\rho \phi, \rho' \phi')$, with $\phi=(\phi_{ij})$ and $\phi'=(\phi'_{ab})$ 
are pairs $(L,\Phi)$ of a quantum channel $\Phi(\rho)=\rho'$, with Choi matrix 
$(S_\Phi)_{\substack{ij \\ ab}}$ and a finite collection
$L = \{ ( L_{\substack{ij \\ ab},r } , (S_{\Phi_r})_{\substack{ij \\ ab}} )\}$
of morphisms $L_{\substack{ij \\ ab},r }: \phi_{ij}) \to \phi'_{ab}$ in $\cA\cC$
with associated $S_{\Phi_r}$ satisfying $\sum_r S_{\Phi_r} =S_\Phi$. 
\end{defn}

 \medskip
 \section{Quantum Information and Gamma Spaces}
 
 As in the case of classical information discussed earlier, we can construct a $\Gamma$-space
 $F_{\cQ\cC}: \Gamma^0 \to \Box_*$ associated to a quantum probabilistic category $\cQ\cC$
 obtained as in the previous section. We can then consider the associated probabilistic
 $\Gamma$-space $F_{\cQ\cC}: \cP\cS_* \to \cP\Box_*$. We consider explicitly the case
 where the underlying category $\cC$ is given by the category of pointed sets $\cS_*$.
 
 \smallskip
 \subsection{Summing functors of quantum pointed sets}
 
 We refer here to the category $\cQ\cC$ with $\cC=\cS_*$ the category
 of finite pointed sets as ``quantum pointed sets". We consider here the
 associated category $\Sigma_{\cQ\cS_*}(X)$ of summing functors 
 $\Theta: P(X) \to \cQ\cS_*$, for a pointed set $(X,\star)$, with $P(X)$
 the category of pointed subsets $(A,\star)$ with morphisms given by
 inclusions.
 
\smallskip

\begin{thm}\label{summingQS}
An object $\Theta$ in the category of summing functors $\Sigma_{\cQ\cS_*}(X)$
is completely specified by the choice of a point 
$\alpha=\{\alpha_x\}_{x\in X\smallsetminus \{ \star \}} \in \cI^N$, with $\# X =N+1$,
and, for each choice of $\alpha$, a set of complex numbers 
$\theta=\{ \theta_x \}_{x\in X\smallsetminus \{ \star \}}$ contained in the annuli
\begin{equation}\label{thetaalpha}
\theta_x \in \cA_x=\{ z\in \C \,:\, \alpha_x (1-\alpha_x) - \frac{1}{4} \leq |z|^2 \leq \alpha_x (1-\alpha_x) \},
\end{equation}
or disks $ \{ |z|^2 \leq \alpha_x (1-\alpha_x) \}$ if $\alpha_x (1-\alpha_x) \leq 1/4$. 
The summing functor then maps $\Theta_{\alpha,\theta}(A)=\rho_A C_A$
where $\rho_A C_A$ consists of a collection of $2^{N_A}\times 2^{N_A}$ pairs
of pointed sets of cardinality $N_A+1=\# A$ with $\rho=(\rho_{ij})$ the
$2^{N_A}\times 2^{N_A}$ density matrix with entries given by the
sequences $(t_1,\ldots, t_{N_A})$ with $t_a\in \{ \alpha_a, 1-\alpha_a, \theta_a, \bar\theta_a\}$
for $a\in A\smallsetminus \{ \star \}$. 
The morphisms of $\Sigma_{\cQ\cS_*}(X)$ are given by the group $\cU(2)^{\otimes N}$
of unitary transformations acting by $U_x \rho^{(x)} U_x^*$ on $$ \rho^{(x)} =
\begin{pmatrix} \alpha_x & \theta_x \\ \bar\theta_x & 1-\alpha_x  \end{pmatrix} $$
and by collections of isomorphisms of pointed sets. 
\end{thm}
 
 \proof Summing functors $\Theta\in \Sigma_{\cQ\cS_*}(X)$ have the properties that  
 $\Theta(\{\star\},\star)=(\{\star\},\star)$ the zero object of $\cQ\cS_*$ 
 and $\Theta(A\cup B)=\Theta(A)\amalg_{\cQ\cS_*} \Theta(B)$ for any
 $A,B\in P(X)$ with $A\cap B=\{ \star \}$. We proceed as in the case of Theorem~\ref{sumsPSstar}.
 The properties of the summing functor implies that it suffices to know the value on sets
 $\Theta(\{ a, \star \})$, since we then obtain
 \begin{equation}\label{ThetaA}
  \Theta(A)=\amalg_{a\in A\smallsetminus \{ \star \}} \Theta(\{ a, \star \}) 
 \end{equation} 
 with the coproduct in $\cQ\cS_*$. As in the case of classical probabilities, 
 we consider $\Theta(\{ a, \star \})$ as a superposition of the two possible choices
of base point $a,\star$ in the set $\{ a, \star \}$, except that now, in addition to the
superposition we also need to account for interference effects. Thus,
we assign to $\{ a,\star\}$ the object in $\cQ\cS_*$ given by the following set
\begin{equation}\label{Thetaastar}
 \Theta(\{ a, \star \})= \left\{ \begin{array}{ll} ( (\{ a,\star\}, \star), (\{ a,\star\}, \star)) & \rho_{11}= \alpha_a  \\
( (\{ a,\star\}, \star), (\{ a,\star\}, a)) & \rho_{12} =\theta_a \\
( (\{ a,\star\}, a), (\{ a,\star\}, \star)) & \rho_{21} =\bar \theta_a \\
( (\{ a,\star\}, a), (\{ a,\star\}, a)) & \rho_{22}= 1-\alpha_a 
\end{array}\right. 
\end{equation}
This reduces to the classical choice in the diagonal case with $\theta_a=0$.
This then determines the value $\Theta(A)$ for all $A\in P(X)$ as the
a list of $2^{N_A}\times 2^{N_A}$ pairs of sets of cardinality $\# A$ with 
associated density matrix $\rho_A = \otimes_{a\in A\smallsetminus \{ \star \}} \rho^{(a)}$, 
the $N_A$-fold tensor product of the matrices as above
\begin{equation}\label{rhoa}
 \rho^{(a)}=\begin{pmatrix} \alpha_a & \theta_a \\ \bar\theta_a & 1-\alpha_a  \end{pmatrix}. 
\end{equation} 
The entries of $\rho_A$ can then be identified with the sequences $(t_1,\ldots, t_{N_A})$ with $t_a\in \{ \alpha_a, 1-\alpha_a, \theta_a, \bar\theta_a\}$ for $a\in A\smallsetminus \{ \star \}$, where the diagonal
entries correspond to those sequences that contain only the letters $\{ \alpha_a, 1-\alpha_a \}$
as in the classical case. 
Thus, in order to specify a summing functor $\Theta$ it suffices to assign
a choice of values $\{ \alpha_x \}_{x\in X\smallsetminus \{ \star \}}$
and of $\{ \theta_a \}_{x\in X\smallsetminus \{ \star \}}$. The only constraint on
the choice of the $\alpha_x$ comes from the normalization of the trace $\Tr(\rho)=1$,
for which, as in the classical case, it suffices to require that all the $\alpha_x\in [0,1]$,
hence $\{ \alpha_x \}_{x\in X\smallsetminus \{ \star \}} \in |\cI^N|$ with $N=\#X-1$.
The constraints on the $\theta_x$ come from the requirement that the density
matrices satisfy $\rho_A \geq 0$. It suffices the ensure that the density matrices $\rho^{(a)}$
of \eqref{rhoa} have non-negative eigenvalues. The characteristic polynomial
$p(\lambda)=\lambda^2 - \Tr(\rho)\lambda + \det(\rho)= \lambda^2-\lambda+\det(\rho)$
has non-negative eigenvalues when the discriminant $\Tr(\rho)^2-4 \det(\rho)=1-4\det(\rho)\geq 0$
and $\det(\rho)\geq 0$. This gives the condition \eqref{thetaalpha}.

The morphisms in $\Sigma_{\cQ\cS_*}(X)$ consist of natural transformations of the functors
that are isomorphisms on objects. This means isomorphisms $\eta_A: \Theta(A)\to \Theta'(A)$
in $\cQ\cS_*$ that are compatible with the inclusions $j: A\hookrightarrow A'$, with
$\eta_A \circ \Theta(j) =\Theta'(j) \circ \eta_A$. An isomorphism $\eta: \rho_A X_A \to \rho_A' X_A'$ in 
$\cQ\cS_*$ consists of an invertible quantum channel mapping $\rho_A$ to $\rho_A'$ and
a collection of isomorophisms of the pairs of pointed sets in the collections $X_A$ and $X_A'$. 
The invertible quantum channel is given by a unitary transformation, and the requirement
that the isomorphisms are natural transformations of the functors, that is, that they
are compatible with the inclusions of subsets, implies that the unitary transformation
$\rho'_A = U_A \rho_A U_A^*$ with $U_A\in \cU(2^{N_A})$ is a product of unitary
transformations of the matrices $\rho^{(a)}$ of \eqref{rhoa}, $U_A=U_{a_1}\otimes \cdots \otimes U_{a_{N_A}}$
with unitaries $U_a\in \cU(2)$. Note that the relation \eqref{thetaalpha} between the off diagonal 
entry $\theta_a$ and the diagonal $\alpha_a$ is preserved under the action of $\cU(2)$. 
\endproof

\smallskip

\begin{prop}\label{cubenerveQC}
The cubical nerve $K=\cN_\fC(\Sigma_{\cQ\cS_*}(X))$ with $K_n={\rm Fun}(\cI^n, \Sigma_{\cQ\cS_*}(X))$
is given by the action groupoid of $\cU(2)^{\otimes N}$ acting on the cubical set 
\begin{equation}\label{ZNset}
 \cZ_N = \bigcup_{Z\in |\cI^N|} \bigcup_{k=0}^N \cI_Z^k \times \cA_k, 
\end{equation} 
with $\cA_k$ a product of $N-k$ annuli (or disks) $\cA_x$ as in \eqref{thetaalpha}.
\end{prop}

\proof A functor $\cI^n \to \Sigma_{\cQ\cS_*}(X)$ assigns to each vertex $v\in \cI^n$ a summing
functor $\Theta_v\in \Sigma_{\cQ\cS_*}(X)$, hence a choice of $\{ \lambda_x, \theta_x \}_{x\in X\smallsetminus \{ \star \}}$ satisfying \eqref{thetaalpha}. Edges of $\cI^n$ correspond to natural transformations between 
$\Theta_v$ and $\Theta_{v'}$ for $\partial(e)=\{ v,v' \}$. Let $\cA_\alpha$ denote as in \eqref{thetaalpha} 
the annulus $\cA_\alpha=\{ \alpha (1-\alpha) - 1/4 \leq |z|^2 \leq  \alpha (1-\alpha) \}$, 
when $\alpha (1-\alpha) > 1/4$. In the case where $\alpha (1-\alpha) \leq 1/4$ we just have the disk 
$\cA_\alpha =\{ |z|^2 \leq  \alpha (1-\alpha) \}$.
Then the datum of a functor 
$\cI^n \to \Sigma_{\cQ\cS_*}(X)$ corresponds to assigning for each vertex $v\in \cI^n$ a union
$\cup_{k=0}^n \P_k(v) \times \cA_k(v)$ of products of
pointed polytopes with $2^k$ vertices and products of annuli (or disks) $\cA_k(v)$, 
where for a given choice of 
$\{ \lambda_x, \theta_x \}_{x\in X\smallsetminus \{ \star \}}$ satisfying \eqref{thetaalpha},
$\P_k(v)$ is the polytope given as in  \eqref{PvProd} by the classical probability space 
$\{0,1 \}^k$ with probability $(t_1 \ldots t_k)$
with $t_i \in \{ \alpha_i (1-\alpha_i) \}$ and $\cA_k(v)$ is a union of products of annuli
$\prod \cA_{\alpha_i}$, considered with their cubical structure. Arguing as in 
Proposition~\ref{cubenerveSum}, the functors associated to adjacent vertices,
hence the corresponding  sequences $t_1,\ldots, t_n$ in the alphabet
$t_i \in \{ \lambda_{x_i}^v,1-\alpha_{x_i}^v \theta_{x_i}^v, \bar\theta_{x_i}^v \}$, 
are related by a morphism in $\Sigma_{\cQ\cS_*}(X)$. These are unitary transformations
in $\cU(2)^{\otimes n}$. In particular, since the sequences $s_1\ldots s_n\in \{ 0,1 \}^n$
labeling adjacent vertices of $\cI^n$ differ at a single digit $s_k$, the corresponding sequences
$t_1,\ldots, t_n$ differ in the action of a single $U_k \in \cU(2)$ relating the density matrices
$\rho^{(x_k)}_v$ and $\rho^{(x_k)}_{v'}$. 
As in Proposition~\ref{cubenerveSum}, this reduces the choices of the data $\{ \alpha^v_x, \theta^v_x \}$
to a single choice $\{ \alpha_x, \theta_x \}_{x\in X\smallsetminus \{ \star \}}$ at a single vertex, with 
the assignments at a the other vertices of the cube obtained by applying unitary transformations
associated to the edges of the cube. Thus, the cubical nerve $\cN_\fC(\Sigma_{\cQ\cS_*}(X))$ can
be described as the action groupoid of the action of $\cU^{\otimes N}$ on the set $\cZ_N$ of
\eqref{ZNset}, which parameterizes the choice of data $\{ \alpha_x, \theta_x \}_{x\in X\smallsetminus \{ \star \}}$. 
\endproof

\smallskip

\begin{lem}\label{PhiXbarC}
The simplicial set $F_{\cQ\cS_*}(X)=\cN_\fC(\Sigma_{\cQ\cS_*}(X))$ is 
homotopy equivalent to the Borel homotopy quotient $\cM_G=EG\times_G \cZ_N$,
with $G=\cU(2)^{\otimes N}$, for $N=\# X-1$, and $\cZ_N$ the set \eqref{ZNset}.
\end{lem}

\proof Proposition~\ref{cubenerveQC} shows that the category $\Sigma_{\cQ\cS_*}(X)$ of
summing functors can be identified with the action groupoid of the group of unitary transformations 
$\cU(2)^{\otimes N}$ acting on density matrices of the form $\rho=\otimes_x \rho^{(x)}$, or
equivalently on the set $\cZ_N$ of \eqref{ZNset} that parameterizes them. 
Thus, the nerve $\cN \Sigma_{\cQ\cS_*}(X)$ can be identified
with the classifying space $B\cG$ of the action groupoid $\cG=\cZ_N \rtimes \cU(2)^{\otimes N}$, 
with $N=\# X-1$.
The classifying space $B\cG$  of an action groupoid $\cG=\cZ \rtimes G$ of a Lie group action
on a manifold is homotopy equivalent to the Borel construction of the homotopy quotient 
$\cZ_G=EG\times_G \cZ$ (see e.g.~Proposition~B.11 of \cite{Watts}).
\endproof

\smallskip

\begin{cor}\label{geomrelQS}
The geometric realization $|\cN_\fC(\Sigma_{\cQ\cS_*}(X))|$ is homotopy
equivalent to a union of strata of the form 
\begin{equation}\label{strataZjP}
 \cZ^j_{N,\cP} \times B\left(\cU(2)^{\otimes j} \otimes (\cU(1)\times \cU(1))^{\otimes (N-j)}\right) 
\end{equation} 
where $\cZ^j_{N,\cP} \subset | \cN_\cC \Sigma_{\cP\cS_*}(X) |$ is a subset of the
realization of $F_{\cP\cS_*}(X)$, the value of the $\Gamma$-space of classical probabilities,
given by
\begin{equation}\label{ZjNP}
\cZ^j_{N,\cP}=\bigcup_{Z\in \cS_j} |\cI^N_Z |,
\end{equation}
with $\cS_j$ the set of sequences $\{ \alpha_x \}$ with $j$ entries equal to $1/2$, 
and where $B(\cU(2)^{\otimes j} \otimes (\cU(1)\times \cU(1))^{\otimes (N-j)})$ are the classifying spaces
of the stabilizers $\cU(2)^{\otimes j} \otimes (\cU(1)\times \cU(1))^{\otimes (N-j)}$ of the strata
$\cZ_N^j$ of $\cZ_N$ with $\cZ^j_{N,\cP}=\cZ^j_N/\cU(2)^{\otimes N}$.
\end{cor}

\proof The homotopy quotient $\cZ_G$ has projection maps
$$ BG \stackrel{\pi_1}{\longleftarrow} \cZ_G \stackrel{\pi_2}{\longrightarrow} \cZ/G, $$
where the projection $\pi_1$ is a fibration over $BG$ with fiber $\cZ$, while the preimage
of a point $x\in \cZ/G$ is a copy of $BG_x$ with $G_x\subset G$ the isotropy group.
In the case of the set $\cZ_N$ of \eqref{ZNset} with the action of $G=\cU(2)^{\otimes N}$,
we can decompose $\cZ_N$ into strata with associated stabilizers of the action. The top
stratum $\cZ_N^0$ consists of 
$\cZ^0_N =\cup_{Z\in \cS_0\subset |\cI^N|} \cup_k \cI_Z^k \times \cA_k$,
where $\cS_0$ is the set of those sequences $Z=\{ \alpha_x \}_{x\in X\smallsetminus \{ \star \}}$ 
where none of the $\alpha_x$ is equal to $1/2$ . The lower strata 
$\cZ_N^j$ consist of $\cZ^0_N =\cup_{Z\in \cS_j\subset |\cI^N|} \cup_k \cI_Z^k \times \cA_k$
where $\cS_j$ is the set of sequences $Z=\{ \alpha_x \}$ where $j$ of the $\alpha_x$ are
equal to $1/2$. Since in the $\cU(2)$ action $\rho^{(x)} \mapsto U_x \rho^{(x)} U_x^*$ 
we can identify unitaries $U_x\in \cU(2)$ up to phase factors (diagonal unitaries) 
$\cU(1)\times \cU(1)$, the stabilizer of the top stratum is $(\cU(1)\times \cU(1))^{\otimes N}$.
The stratum $\cZ^j_N$ has stabilizer $\cU(2)^{\otimes j} \otimes (\cU(1)\times \cU(1))^{\otimes (N-j)}$. 
The quotient $\cZ_N/\cU(2)^{\otimes N}$ is correspondingly decomposed into strata,
where up to the action of $\cU(2)$ we can identify the density matrices $\rho^{(x)}$
with diagonal classical probabilities. This implies that we can identify the quotient
$\cZ_N/\cU(2)^{\otimes N}$ with a classical space 
$$  | \cN_\cC \Sigma_{\cP\cS_*}(X) | = \bigcup_{Z\in |\cI^N|} |\cI^N_Z | $$
with a stratification by \eqref{ZjNP}. Over each set 
$\cZ^j_{N,\cP}$ we have a copy of the fiber $B(\cU(2)^{\otimes j} \otimes (\cU(1)\times \cU(1))^{\otimes (N-j)}$.
\endproof

\smallskip

\begin{rem}\label{topgeomrel}{\rm 
The geometric realization of the nerve $\cN_\fC(\Sigma_{\cQ\cS_*}(X))$ is
more interesting topologically than the case 
of classical probabilities, due to the presence of the classifying spaces 
of unitary groups.}
\end{rem}

\smallskip

\begin{rem}\label{SpecQS}{\rm 
We obtain a $\Gamma$-space of quantum pointed sets
$F_{\cQ\cS_*}: \Gamma^0 \to \Box_*$ that assigns to
a pointed set $(X,\star)$ the cubical nerve $\cN_\fC(\Sigma_{\cQ\cS_*}(X))$
of the category of summing functors of Theorem~\ref{summingQS}.
This can be extended to an endofunctor $F_{\cQ\cS_*}: \Box_*\to \Box_*$
and determines an associated homotopy theoretic spectrum,
by the Segal construction \cite{Segal}. }
\end{rem}

\smallskip
\subsection{The stochastic Gamma spaces}

As in the case of classical probabilities, we can associate to a
category $\cQ\cC$ a probabilistic
$\Gamma$-space $F_{\cQ\cC}: \cP\cS_* \to \cP\Box_*$. The
construction is analogous to the case of $F_{\cP\cC}: \cP\cS_* \to \cP\Box_*$
that we discussed earlier. By  Lemma~\ref{nervePC} the functor
$F_{\cQ\cC}$ maps a probabilistic pointed set $\Lambda X$ to 
the cubical nerve $\cN_\fC( \cP\Sigma_{\cQ\cC}(\Lambda X) )$,
which can be identified with the probabilistic pointed cubical set 
$\sum_i \lambda_i \cN_\fC(\Sigma_{\cP\cQ}(X_i,x_i))$. 

\medskip
\section{Gamma spaces, spectra, and gapped systems}

There has been a growing interest recently in the use of homotopy theoretic
methods in the modeling of topological phases of matter. In particular, the
use of spectra to classify symmetry protected topological phases in terms
of generalized cohomology groups associated to
a spectrum of invertible physical systems, \cite{Kitaev}.
This idea is elaborated upon in \cite{Gaio}, where a spectrum ${\rm GP}^\times_n$ of 
invertible gapped phases of matter is considered, with homotopy equivalences 
$\Omega {\rm GP}^\times_n \to {\rm GP}^\times_{n-1}$ corresponding to
realizing a continuous transition between $n$-dimensional systems in the same 
phase via an invertible $(n-1)$-dimensional interface (invertible defect). 

\smallskip

We consider here a similar idea, from the point of view of Segal's $\Gamma$-spaces
and we present a construction of $\Gamma$-spaces associated to gapped systems.
Our setting here is only a simplified model of the properties one usually
requires for gapped systems, see \S 5.2 of \cite{ZCZW}. In general, in addition
to the existence of a gap $\Delta>0$ in the spectrum, one also requires a
uniform bound on the degeneracy of the ground state, namely the
condition that for all $(X,H_X)$ the ground state degeneracy 
satisfies $1\leq \dim\Ker(H_X) \leq m$ with some uniform bound by some 
fixed $m \in \N$. In the setting we consider here,
this condition would not be compatible with the categorical sum. 
In order to obtain a more sophisticated model for
gapped systems where the uniformly bounded degeneracy condition can
also be imposed, we need to work with a different categorical setting. 
This will be investigated elsewhere.

\smallskip
\subsection{Gamma spaces of gapped systems}

In this setting, instead of considering the category $\cF\cQ$ of
finite quantum probabilities (density matrices) with morphisms
given by quantum channels, one considers a category $\cF\cQ^\Delta$
where the objects are systems $(X,H_X)$ with a Hamiltonian $H_X$
acting on a HIlbert space $\cH_X=\oplus_{x\in X}\cV_x$ with a fixed
internal space $\cV_x$ (which for simplicity we will just take equal to a
line $\C_x$), with the property that $H_X^*=H_X$ and that
$H_X$ has a gap in the spectrum above the ground level,
that is, $0\in \Sp(H_X)$ and $\Sp(H_X)\subset
\{ 0 \} \cup [\Delta,\infty)$. We realize objects in $\cF\cQ^\Delta$
as objects of $\cF\cQ$ by associating to a pair $(X, H_X)$ the
pair $(X,\rho_X)$ in $\cF\cQ$ with 
$$ \rho_X = \frac{e^{-\beta H_X}}{\Tr(e^{-\beta H_X})}, $$
where $\beta>0$ is a fixed inverse temperature parameter. 
Using this identification of objects of $\cF\cQ^\Delta$ with
a subset of objects of $\cF\cQ$, we take the 
morphisms in $\cF\cQ^\Delta$ to be induced by the morphisms in $\cF\cQ$.
Namely, morphisms in ${\rm Mor}_{\cF\cQ^\Delta}((X,H_X),(Y,H_Y))$ 
are gap preserving quantum channels, $\Phi(\rho_X)=\rho_Y$.

\smallskip

When we apply the previous construction of quantum categories $\cQ\cC$
using the category $\cF\cQ^\Delta$ instead of $\cF\cQ$, we obtain a
subcategory $\cQ\cC^\Delta$ of $\cQ\cC$, which is
described as follows.

\smallskip

\begin{defn}\label{GappedQS}
The gapped quantum category $\cQ\cC^\Delta$ has objects
that include the zero object of $\cQ\cC$ and objects 
$\rho C=((C_i,C_j), \rho_{ij})$ of $\cQ\cC$ where the density matrix
$\rho$ is of the form
\begin{equation}\label{rhoH}
\rho = \frac{e^{-\beta H}}{\Tr(e^{-\beta H})},
\end{equation}
for some (fixed) inverse temperature parameter $\beta>0$, with the property that the
Hamiltonian $H$ has spectrum $\Sp(H)\subset \{ 0 \}\cup [\Delta,\infty)$, 
for a fixed gap $\Delta>0$. The morphisms in $\cQ\cC^\Delta$ are induced by 
the morphisms in $\cQ\cC$.
\end{defn}

\smallskip

\begin{lem}\label{coprodDelta}
The coproduct in $\cQ\cC$ induces a coproduct in $\cQ\cC^\Delta$.
\end{lem}

\proof The coproduct $\rho C \amalg_{\cQ\cC} \rho' C'$ of two objects
$\rho C, \rho' C'\in \cQ\cC^\Delta$ has density matrix given by the
product $\rho \otimes \rho'$. We have
$\rho=e^{-\beta H}/\Tr(e^{-\beta H})$ and $\rho'=e^{-\beta H'}/\Tr(e^{-\beta H'})$
where the respective Hamiltonians
$H, H'$ have spectrum contained in $\{ 0 \} \cup [\Delta,\infty)$. The tensor
product $\rho\otimes \rho'$ corresponds to the Kronecker sum $H\oplus H'=H \otimes 1 + 1 \otimes H'$
of the Hamiltonians. The spectrum of the Kronecker sum $\Sp(H \otimes 1 + 1 \otimes H')$
is given by sums of eigenvalues of $H$ and $H'$, hence it is still contained in $\{ 0 \} \cup [\Delta, \infty)$,
so that the coproduct in $\cQ\cC$ of two objects in $\cQ\cC^\Delta$ is still an object in $\cQ\cC^\Delta$.
\endproof

\smallskip

We focus on the case where $\cC=\cS_*$, the category of finite pointed sets.
As in the case of $\cQ\cS_*$, we construct the associated $\Gamma$-space
by constructing the category $\Sigma_{\cQ\cS_*^\Delta}(X)$ of summing
functors $\Theta: P(X)\to \cQ\cS_*^\Delta$.

\smallskip

\begin{prop}\label{summingQSDelta}
For sufficiently large $\beta>0$, 
an object $\Theta$ in $\Sigma_{\cQ\cS_*^\Delta}(X)$ is specified by the choice
of a point $\alpha=\{ \alpha_x \}_{x\in X\smallsetminus \{ \star \}}\in \cI_{\beta,\Delta}^N$ 
for an interval $\cI_{\beta,\Delta}=[a_{\beta,\Delta}, b_{\beta,\Delta}]\subset [0,1]$ and with $N=\# X-1$,
and a choice of $\{ \theta_x \}_{x\in X\smallsetminus \{ \star \}} \in T^N_{r(Z)}$, where $T^N=(S^1)^N$ 
is a torus and the subscript $r(Z)$ indicates that the $k$-th circle has a radius $r=r(\alpha_x,\beta,\Delta)$
uniquely determined by the choice of $\alpha_x$ and by the fixed values of $\Delta$ and $\beta$.
The morphisms in $\Sigma_{\cQ\cS_*^\Delta}(X)$ are given 
by unitary transformations in $\cU(2)^{\otimes N}$  
and by collections of isomorphisms of pointed sets.
\end{prop}

\proof As in the case of $\cQ\cS_*^\Delta$, we know that the values $\Theta(A)$
are given by coproducts \eqref{ThetaA} in $\cQ\cS_*^\Delta$, with the terms
$\Theta(\{ a, \star \})$ as in \eqref{Thetaastar}. In this case, the 
density matrix  \eqref{rhoa} associated to $\Theta(\{ a, \star \})$ will have to
satisfy additional constraints due to the gap condition on the spectrum of
the associated Hamiltonian. In the case of a $2\times 2$ matrix, the condition
that the spectrum has a gap of width $\Delta$ above the ground level $\lambda=0$
corresponds to requiring that one of the eigenvalues is zero and the other one
is equal to the width of the gap $\Delta$. This means that the spectrum of the
corresponding density matrix $\rho=e^{-\beta H}/\Tr(e^{-\beta H})$ is given by
$$ \Sp(\rho)=\{ \frac{e^{-\beta \Delta}}{1+e^{-\beta \Delta}}, \frac{1}{1+e^{-\beta \Delta}} \}. $$
Let $q=1-\alpha(1-\alpha)+|\theta|^2$. Then the condition above on the eigenvalues of \eqref{rhoa} gives
$$ \frac{1}{2} (1-q^{1/2}) = \frac{e^{-\beta \Delta}}{1+e^{-\beta \Delta}}, \ \ \ \  \frac{1}{2} (1+q^{1/2}) =
\frac{1}{1+e^{-\beta \Delta}}, $$
which gives 
$$ q^{1/2} = \frac{1-e^{-\beta \Delta}}{1+e^{-\beta \Delta}}. $$
This then gives the relation
\begin{equation}\label{thetaalphaDelta}
 |\theta|^2 = \left(\frac{1-e^{-\beta \Delta}}{1+e^{-\beta \Delta}}\right)^2 -1 + \alpha(1-\alpha)
= \frac{-4e^{-\beta\Delta}}{(1+e^{-\beta \Delta})^2} + \alpha(1-\alpha), 
\end{equation}
where $0< 4e^{-\beta\Delta}/(1+e^{-\beta\Delta})^2 \leq 1$. There is an interval of values
$0< e^{-\beta \Delta} \leq u_{\beta,\Delta}$ with $u_{\beta,\Delta}<1$ such that the
discriminant of $\frac{-4e^{-\beta\Delta}}{(1+e^{-\beta \Delta})^2} + \alpha(1-\alpha)=0$,
seen as an equation in $\alpha$, is non-negative. Then the 
right-hand-side of \eqref{thetaalphaDelta} is non-negative for $\alpha$ in the interval
$[a_{\beta,\Delta}, b_{\beta,\Delta}]$ between the two roots. 
For a fixed value of the gap $\Delta$, it is always possible to choose an inverse
temperature $\beta >0$ sufficiently large so that the condition $e^{-\beta \Delta} \leq u_{\beta,\Delta}$
is satisfied. For such a choice of $\beta$, one then obtains solutions of \eqref{thetaalphaDelta}
given by any choice of $\alpha \in [a_{\beta,\Delta}, b_{\beta,\Delta}]$ and a circle of values
of $\theta$ with radius $r=r(\alpha,\Delta,\beta)$ 
fixed by the relation \eqref{thetaalphaDelta} (and depending on $\alpha$
and on $\Delta$ and $\beta$).
The morphisms in in $\Sigma_{\cQ\cS_*^\Delta}(X)$ are given 
by unitary transformations in $\cU(2)^{\otimes N}$
acting by $U_x \rho^{(x)} U_x^*$ on the density matrices $\rho^{(x)}$ 
as in Theorem~\ref{summingQS}, and by collections of isomorphisms 
between the pointed sets in the two objects. Since the locus determined
by the relation \eqref{thetaalphaDelta} is specified by the condition
on the spectrum of the matrices $\rho^{(x)}$, it is preserved by 
unitary transformations. 
\endproof

\smallskip

\begin{prop}\label{geomrealDelta}
The nerve $\cN_\fC(\Sigma_{\cQ\cS_*^\Delta}(X))$ is the action groupoid of the
$\cU(2)^{\otimes N}$ action on the cubical set
\begin{equation}\label{ZNDelta}
\cZ_{N,\Delta} = \bigcup_{Z\in |\cI_{\beta,\Delta}^N|}\,\, \bigcup_{k=0}^N \cI_Z^k \times T_{r(Z)}^{N-k}.
\end{equation}
The geometric realization $|\cN_\fC(\Sigma_{\cQ\cS_*^\Delta}(X))|$ is homotopy
equivalent to a union of strata of the form \eqref{strataZjP} where the 
$\cS_j \subset |\cI_{\beta,\Delta}^N|$
consists of all the sequences $\{ \alpha_x \} \in [a_{\beta,\Delta}, b_{\beta,\Delta}]^N$ 
where $j$ of the terms are equal to $1/2$. 
\end{prop}

\proof The argument is analogous to Proposition~\ref{cubenerveQC} 
and Corollary~\ref{geomrelQS}. 
The nerve $\cN_\fC (\Sigma_{\cQ\cS_*^\Delta}(X))$ is constructed as in 
the case of $\cQ\cS_*$, except that in this case the annuli and disks are replaced
by circles $\theta_x\in S^1_{r(\alpha,\Delta,\beta)}$ of radius determined by 
\eqref{thetaalphaDelta}. We write $T^{N-k}_{r(Z)}$ for the product of these
$N-k$ circles, where we write $r(Z)$ for this dependence of the radii on the 
$\alpha_x$, leaving the dependence on $\beta$ and $\Delta$ implicit.
For fixed $\Delta$, we are choosing $\beta>0$ large enough as in Proposition~\ref{summingQSDelta},
so that for $\alpha$ in the subinterval $[a_{\beta,\Delta}, b_{\beta,\Delta}]\subset [0,1]$
the estimate $\frac{-4e^{-\beta\Delta}}{(1+e^{-\beta \Delta})^2} + \alpha(1-\alpha)\geq 0$ holds.
The interval $[a_{\beta,\Delta}, b_{\beta,\Delta}]$ contains the point $\alpha=1/2$ as one
can verify directly. 
\endproof

\smallskip

\begin{rem}\label{SpecDelta}{\rm 
The $\Gamma$-space $F_{\cQ\cS_*^\Delta} : \Gamma^0 \to \Box_*$ obtained
in this way can be extended to an endofunctor $F_{\cQ\cS_*^\Delta} : \Box_* \to \Box_*$
and determined an associated connective spectrum by the Segal construction of
\cite{Segal}. This provides then a construction of a homotopy theoretic spectrum
associated to a category of gapped systems with a fixed gap $\Delta$. 
}\end{rem}
 
\smallskip

\begin{rem}\label{probGammaDelta}{\rm 
As in the cases of $\cP\cS_*$ and of $\cQ\cS_*$ it is also possible to
extend the $\Gamma$-space $F_{\cQ\cS_*^\Delta} : \Gamma^0 \to \Box_*$ 
to a probabilistic $\Gamma$-space $F_{\cQ\cS_*^\Delta} :\cP\cS_* \to \cP\Box_*$.
}\end{rem}

\smallskip
\subsection{Gamma spaces and gapped phases}

We consider here a different construction of a $\Gamma$-space related to gapped
system, where instead of fixing the gap and restricting the category $\cF\cQ$ to
a subcategory $\cF\cQ^\Delta$ of gapped systems and gap preserving quantum
channels, we consider all the objects of $\cF\cQ$, so that there is no fixed gap,
but we change the morphisms so that we regard all the quantum channels that
preserve a gap $\Delta>0$ as isomorphisms. We can do this in the form of a
localization of the category $\cF\cQ$ at a collection of morphisms $\cT_\Delta$. 

\smallskip

More precisely, the set of morphisms $\cT_\Delta$ consists
of all morphisms in $\cQ\cS_*$ where both source and target
are objects in $\cQ\cS_*^\Delta$. The localization $\cQ\cS_*[\cT^{-1}_\Delta]$ is obtained as
a quotient of the path category $\cP(\cQ\cS_*, \cT^{-1}_\Delta)$.
The path category has the same objects as $\cQ\cS_*$
and morphisms given by arbitrary concatenations
$\Psi_1\cdots \Psi_N$ where the $\Psi_i$ are either
morphisms in $\cQ\cS_*$ or formal inverses of morphisms
in $\cT_\Delta$, with the target of $\Psi_i$ equal to the
source of $\Psi_{i+1}$. The equivalence relation on $\cP(\cQ\cS_*, \cT^{-1}_\Delta)$
identifies the empty string at a given object with the identity morphism,
a string $\Psi_1 \Psi_2$ where both $\Psi_i$ are morphisms in $\cQ\cS_*$
with the morphism $\Psi_2\circ \Psi_1$ and a string $\Phi^{-1} \Phi$
or $\Phi \Phi^{-1}$, with $\Phi\in \cT_\Delta$ and $\Phi^{-1}$
its formal inverse, with the identity morphism on the source,
respectively target, of $\Phi$. This determines a category
$$ \cQ\cS_*[\cT^{-1}_\Delta]=\cP(\cQ\cS_*, \cT^{-1}_\Delta)/\sim $$
with a localization functor $\cQ\cS_* \to \cQ\cS_*[\cT^{-1}_\Delta]$
which maps morphisms in $\cS_\Delta$ to isomorphisms.
We refer the reader to the overview in \S~2 of \cite{Fritz2}. 

\smallskip

The category $\cQ\cS_*[\cT^{-1}_\Delta]$ has zero object and
categorical sum inherited from $\cQ\cS_*$.  Thus, we can
consider the $\Gamma$-space 
$F_{\cQ\cS_*[\cT^{-1}_\Delta]}: \Gamma^0\to \Box_*$
associated to $\cQ\cS_*[\cT^{-1}_\Delta]$.

\smallskip

\begin{prop}\label{GammaDeltaLoc}
The summing functors $\Theta: P(X) \to \cQ\cS_*[\cT^{-1}_\Delta]$ are specifed
by data $\{ \alpha_x \}_{x\in X\smallsetminus\{ \star \}} \in |\cI^N|$ and
$\{ \theta_x \} \in \cA_x$ as in the case of $\cQ\cS_*$. 
For sufficiently large $\beta>0$, the morphisms in $\Sigma_{\cQ\cS_*[\cT^{-1}_\Delta]}(X)$ are
given by unitary transformations in $\cU(2)^{\otimes N}$ and by 
quantum channels $\Phi^{(x)}$ with $\Phi^{(x)}\rho^{(x)}=\rho^{\prime\, (x)}$
whenever both $\rho^{(x)}$ and $\rho^{\prime\, (x)}$ have entries satisfying
the relation \eqref{thetaalphaDelta}.
\end{prop}

\proof The characterization of objects in $\Sigma_{\cQ\cS_*[\cT^{-1}_\Delta]}(X)$ is
as in the case of the category $\cQ\cS_*$. 
Morphisms in the category $\Sigma_{\cQ\cS_*[\cT^{-1}_\Delta]}(X)$ of summing functors
are isomorphisms in $\cQ\cS_*[\cT^{-1}_\Delta]$ compatible with the inclusions
of subsets in $P(X)$. These isomorphisms are generated by unitary transformations
$\cU(2)^{\otimes N}$ and by morphisms in $\cT_\Delta$, 
and by isomorphisms of pointed sets. A morphism in $\cT_\Delta$ compatible with
the inclusions can be expressed in terms of quantum channels relating the
density matrices $\rho^{(x)}$ of the source and target object where both source
and target are in $\cQ\cS_*^\Delta$. 
\endproof  

\smallskip

\begin{cor}\label{nerveTDelta}
The nerve $\cN_\fC(\Sigma_{\cQ\cS_*[\cT^{-1}_\Delta]}(X))$ is given by the
homotopy quotient of the equivalence relation on the set $\cZ_N$ of \eqref{ZNset}
generated by quantum channels that preserve the $\Delta$-gap in the spectrum.
\end{cor}

\proof By the same argument used in the case of the category $\cQ\cS_*$,
we identify the cubical nerve with the groupoid of the equivalence relation on
$\cZ_N$ given by unitary equivalence implemented by the group $\cU(2)^{\otimes N}$
together with the action of the gap preserving quantum channels in $\cT_\Delta$.
\endproof

\smallskip

\begin{rem}\label{samephase}{\rm
Equivalent objects in $\cN_\fC(\Sigma_{\cQ\cS_*[\cT^{-1}_\Delta]}(X))$ 
under unitaries in $\cU(2)^{\otimes N}$ and quantum channels in $\cT_\Delta$ 
corresponds to system in the same topological phase,
in the sense that they are mapped to one another in such a way that
the gap $\Delta$ is preserved.
}\end{rem}
  
 \smallskip

The spectra associated to the $\Gamma$-spaces 
$\cF_{\cQ\cS_*^\Delta}: \Gamma^0\to \Box_*$ and
$\cF_{\cQ\cS_*[\cT^{-1}_\Delta]}: \Gamma^0\to \Box_*$
are obtained by considering the corresponding extensions
to endofunctors of $\Box_*$,
$$ F_{\cC}(K)= \int^n K_n \wedge F_{\cC}(\{0,\ldots,n\}) $$
for $\cC$ equal to either $\cQ\cS_*^\Delta$ or $\cQ\cS_*[\cT^{-1}_\Delta]$.
In the first case, the cubical set
$\cF_{\cQ\cS_*^\Delta}(K)$, for an $n$-dimensional cubical set $K$,
describes the homotopy quotient of the unitary equivalence relation
on systems on the $n$-dimensional $K$ with gap $\Delta$,
while in the second case it describes the homotopy quotient of the
equivalence relation by gap preserving quantum channels on all
systems on the $n$-dimensional sets. The structure maps
$S^1\wedge F_{\cC}(S^n) \to F_{\cC}(S^{n+1})$ relate these
equivalences of systems in different dimensions via a suspension
operation. 
 
 \bigskip
 
 \subsection*{Acknowledgment} The author is extremely grateful to Tobias Fritz
 for many useful discussions and for providing many comments and suggestions.
 She also thanks Paolo Aluffi, Tom Leinster, and Jack Morava for helpful comments.
 The author is partially supported by NSF grant 
 DMS-1707882, and by NSERC Discovery Grant RGPIN-2018-04937 and Accelerator
Supplement grant RGPAS-2018-522593, and by the Perimeter Institute for Theoretical Physics.

\end{document}